\documentclass[usegraphicx,usenatbib]{mn2e}
\usepackage{amssymb}
\usepackage{graphicx}

\def\Teff{$T_{\rm eff}$}
\def\logg{$\log\,g$}

\def\Vt{V${\rm t}$}

\def\aj{AJ}%
\def\araa{ARA\&A}%
\def\apj{ApJ}%
\def\apjl{ApJ}%
\def\apjs{ApJS}%
\def\ao{Appl.~Opt.}%
\def\aap{A\&A}%
\def\aaps{A\&AS}%
\def\mnras{MNRAS}%
\def\pasj{PASJ}%
\def\nat{Nature}%
\newcommand {\apgt} {\ {\raise-.5ex\hbox{$\buildrel>\over\sim$}}\ }
\newcommand {\aplt} {\ {\raise-.5ex\hbox{$\buildrel<\over\sim$}}\ }

\title[Behaviour of elements ]
{Behaviour of elements from lithium to europium in stars
with and without planets
\thanks{Tables A1-A3 are only available in electronic form}
}
\author[T.~Mishenina  et al.]
{T.~Mishenina$^{1,2}$,
 V.~Kovtyukh$^{1,2}$,
C.~Soubiran$^{3}$,
 V.Zh.~Adibekyan$^{4}$\\
$^{1}$Astronomical Observatory, Odessa National University,        
           Shevchenko Park, UA-65014, Odessa, Ukraine, {\it e-mail: tmishenina@ukr.net}\\
$^{2}$ Isaac Newton Institute of Chile, 
 Odessa Branch, Shevchenko Park, UA-65014 Odessa, Ukraine\\
$^{3}$ Universit\'e de Bordeaux 1 - CNRS - Laboratoire d'Astrophysique de Bordeaux,\\
UMR 5804, BP 89, 33271 Floirac Cedex, France  \\
$^{4}$ Instituto de Astrof\'isica e Ci\^encias do Espa\c{c}o, Universidade do Porto, CAUP, 
Rua das Estrelas, 4150-762 Porto, Portugal \\
}

\begin{document}

\date{Accepted 2016 xxx. Received 2016 xxx; in original form 2016 xxx}
\pagerange{\pageref{firstpage}--\pageref{lastpage}}
\pubyear{2016}

\maketitle

\label{firstpage}

\begin{abstract}
We conducted an analysis of the distribution of elements from lithium to europium in
200 dwarfs in the solar neighbourhood ($\sim$  20 pc) with temperatures in
the range 4800--6200 K  and metallicities [Fe/H] higher than --0.5 dex.
Determinations of atmospheric parameters and the chemical
composition of the dwarfs were taken from our previous studies.
We found that the lithium abundances in the planet-hosting solar-analog 
stars of our sample were lower than those in the stars without planetary systems.
Our results reveal no significant differences exceeding the determination
errors for the abundances of investigated elements, 
except for aluminium and barium,  which are more and less abundant in the planet-hosting stars, 
respectively. We did not find  confident dependences  of the lithium, 
aluminium and barium abundances on the ages of our target stars  
(which is probable because of the small number of stars). Furthermore, we  
found no correlation between the abundance differences in [El/Fe] and the condensation
temperature ($T_{cond}$) for stars in the 16 Cyg binary system, unlike the case for
51 Peg (HD 217014), for which a slight excess of volatile elements
and a deficit of refractories were obtained relative to those of solar twins. 
We  found that one of the components of 16 Cyg exhibits a slightly  higher average abundance 
than its counterpart ($<$[El/H](A--B)$>$ = 0.08 $\pm$0.02 dex); however, no significant abundance 
trend versus $T_{cond}$ was observed. Owing to the relatively large errors, we cannot provide 
further constraints for this system.
\end{abstract}


\section{Introduction}
A large number of studies of planet-hosting stars 
\citep[][etc]{gonzalez:97, gonzalez:98,  santos:00, fischer:05, udry:07,
adibekyan:12a, adibekyan:12b} have been carried out 
over recent decades with the aim of increasing our 
understanding of the processes involved in the formation of planets.
The earliest studies of the metallicity of
planet-hosting  stars  \cite[e.g.][]{gonzalez:97}, as well as subsequent  ones 
\citep[see][and reference therein]{udry:07} indicated that most of these stars are
 rich in metals. There are two possible explanations for the excess 
metallicity observed. The first is the infall of metal-rich (planetary) material into 
the stellar envelope 
\citep[e.g.][]{gonzalez:98}  and the second is related to the 
prestellar enrichment of interstellar matter  \citep[][and references therein]{fischer:05}. 
The issue has not yet been finally resolved; however, the
second assumption is to be preferred, as the infall of matter
cannot provide an appreciable increase in the metal abundance.

The high metallicity of planet-hosting stars
 is well-established for 
stars with massive planets (Jupiter-like ones)
\citep[][]{gonzalez:97, santos:00, santos:01, fischer:05, sousa:08},
while it is not for stars with less
massive planets, namely those with planets the size of Neptune or
Earth \citep[e.g.][]{udry:07, sousa:08, sousa:11, wang:15}.
The results of \cite{buchhave:12} suggest that terrestrial planets
have  no special requirement 
for enhanced metallicity for their formation, and they support the hypothesis
that stars hosting terrestrial planets have a metallicity similar 
to stars with no known transiting planets \citep{buchhave:15}.
\cite{sousa:08} and \citep{adibekyan:12a, adibekyan:12b}
drew attention to the possibility of planet formation for metallicities lower
than solar. It was shown that there is an excess of
$\alpha$-elements, especially magnesium, in the stars with
low-mass planets, which is more pronounced in the thick-disc population 
than in the thin disc for metallicities below --0.3 dex \citep{adibekyan:12a, adibekyan:12b}.
This implies that metals other than iron may  
noticeably contribute to planetary formation if the iron abundance is low.

Lithium plays a unique role in the study of planet-hosting
stars. Its abundance is apparently lower in the stars with planetary systems than
in those without   
 \citep[][]{gonzalez:00, gonzalez:08, gonzalez:10, israelian:04,  israelian:09, 
delgado:14,  figueira:14, delgado:15}. In particular, \cite{delgado:14}
found that there is some evidence that lithium depletion  in planet-hosting solar-type 
stars is higher when their planets are more massive than Jupiter.
Hot stars that host Jupiter-like planets and have effective temperature \Teff\ in 
the range of 5900-6300 K 
show  lithium abundances that are 0.14 dex lower than those in stars without detected planets
\citep{delgado:15}.
It should be noted, however, that lithium depletion is usually
associated with stellar evolution
\citep[][]{deniss:10, talon:05, anrassy:15}, a phenomenon
that is confirmed by the
dependence of lithium abundances on age
\citep[e.g.][]{monroe:13, melendez:14, carlos:16}. \cite{carlos:16} 
found that the lithium abundances of solar twin stars are a function
of stellar age, while there is no indication of any  relationship between planet-hosting stars
and enhanced lithium depletion. 

In order to identify either the presence or the absence of a possible relation between chemical abundances and mechanisms of planetary formation, various studies have been performed 
to examine the chemical peculiarities of planet-hosting stars
\citep[e.g.][]{melendez:09, ramirez:09, adibekyan:15b}, the  
main properties of planets and their hosts (e.g. mass)
\citep{kang:11, dorn:15, sousa:15},  
their position in the Galaxy \citep{haywood:08, haywood:09, adibekyan:14}, and
Galactic evolution \citep[e.g.][]{adibekyan:15b}.  
Studies of the chemical composition of the Sun, solar twins and solar analogues that have
 highly accurate abundance determinations ($\sim$ 0.01 dex), \cite{melendez:09} 
shown a decrease in the relative content of refractory elements in the Sun, namely one that is 
by 20$\%$ lower than those of solar-analogue stars and solar twins with giant planets. 
Because a correlation between the abundance differences and condensation temperature $T_{cond}$ 
was found, these authors speculated that the decrease in the relative content was associated 
with the presence of Earth-like planets.  
In the literature, there are several alternative explanations 
for the abundance trends with $T_{cond}$. \cite{adibekyan:14} found that chemical peculiarities 
(i.e. small refractory-to-volatile ratio) of planet-hosting stars are likely to reflect their older age 
and inner Galactic origin; hence, stellar age and, probably, Galactic origin are the key factors  
to establish the abundances of some specific elements. It has also been suggested that the $T_{cond}$ trend correlates strongly with stellar radius and mass \citep{mald:15, mald:16}. The trend may also 
depend on stellar environment \citep{onehag:14}, and internal processes, such as gas-dust 
segregation in the protostellar disc \citep{gaidos:15}. 
Recently, \cite{adibekyan:16} found that the $T_{cond}$ trend may depend strongly 
 on the spectra of the stars used. In particular, these authors observed 
significant differences in the abundances of the same star as derived from different high-quality spectra.
Previously, \cite{sousa:08}  had suggested that the detectability
of Neptune-class planets may increase in stars with a low metallicity. 
\cite{kang:11} also confirmed  the
presence of chemical abundance differences between stars with and
without exoplanets, as well as the relationship between chemical abundances and
planetary mass.
\cite{sousa:15} studied the effect of stellar mass on the derived
planetary mass and noted that the stellar mass estimates for giant stars should be
employed with extreme caution when computing planetary masses.

Binary systems of stellar twins are important objects in this investigation,  
as the effects of 
stellar age (chemical evolution) or birthplace in the Galaxy are similar for  both components  of 
the binary pair. \cite{nissen:15} showed  that there are clear correlations between [El/Fe] ratios 
and stellar ages in solar twins.  \cite{teske:16}  found that  both components in the binary 
system WASP 94 A and B are planet-hosting stars, and they differ in their chemical composition. 
The binary system HD 80606/HD80607 exhibits no remarkable differences in the abundances of its components, despite 
the fact that the star HD 80606 hosts a giant planet \citep{saffe:15, mack:16}. However,
abundance variations were found in the XO-2 planet-hosting binary from independent 
observations \citep{biazzo:15, ramirez:15, teske:15}. As reported by \cite{spina:16} 
the abundance ratios [El/Fe] show signatures of both chemical evolution and planets.

Our group has investigated nearly 600 stars, dwarfs and giants over
several years. We determined their atmospheric parameters and chemical
composition in order to study the chemical and dynamical evolution of the Galaxy and,
primarily, the chemical enrichment of various Galactic substructures
\citep[][]{mishenina:04, mishenina:06, mishenina:08, mishenina:12,
mishenina:13, mishenina:15a, mishenina:15b}. We detected 14 planet-hosting stars among 
the target ones.
It would be interesting to examine the behaviour of the abundances of various
elements, including lithium, refractory and volatile elements, as well
as neutron-capture elements for the stars with and without planets that
are present in our data base. Such a study could enable an independent analysis of 
the correlation between the presence of planets and the chemical composition of stars,
possibly shedding some light on the existing contradictions, as well
as re-examining earlier determined constraints on the mechanisms
of formation of planets and planetary systems.

The paper is set out as follows.
The observations and selection of stars, as  well as determination of stellar
parameters and elemental abundance are described in \S \ref{sec: stellar param}.
The analysis of element abundances is presented in
\S \ref{sec: el_analys}. In \S \ref{sec: Tcond_abun}, the
correlations between elemental abundances and condensation temperatures are discussed.
The connection with chemical enrichment of galactic disc
is reported in \S \ref{sec: connect_struct}.
\ref{sec: conclus} summarizes and concludes the paper.

\section{Observations and determination of parameters and chemical composition}
\label{sec: stellar param}

Our sample consists of 200 dwarfs with temperatures in the range
4800 -- 6200 K and metallicities
from --0.4 dex that correspond to the range of stellar parameters of
14 stars that host planets  
(Exoplanets.eu database; \cite{schneider:11}) from our data base.
Table \ref{par_pl} lists the main characteristics of the
planet-hosting stars and planets
(with their masses  in units of the mass of Jupiter (M$_J$)).

\begin{table*}
\caption{Characteristics of the planet-hosting stars and planets. Atmospheric parameters 
are from our previous papers
(Mishenina et al. 2004, 2008, 2012) while the masses of the
planets are from the Exoplanets.eu data base (Schneider et al. 2011).}
\label{par_pl}
\begin{tabular}{rccrll}
\hline
HD   &   \Teff (K) &\logg  & [Fe/H] &  Planet   &      Mass (M$_J$)   \\
\hline
3651  & 5277$\pm$ 5 & 4.5$\pm$0.20 &   0.15$\pm$0.10 &   HD 3651 b&    0.2      \\
7924  & 5165$\pm$ 5 & 4.4$\pm$0.20 & --0.22$\pm$0.10 &   HD 7924 d&    0.0203   \\
      &              &             &                &   HD 7924 c&    0.0247   \\
      &              &             &                &   HD 7924 b&    0.0273   \\
9826  & 6074$\pm$ 10 & 4.0$\pm$0.20 &   0.10$\pm$0.10 &   ups And c&    1.8      \\
      &              &             &                &   ups And d&   10.19     \\
      &              &             &                &   ups And b&    0.62     \\
38858 & 5776$\pm$ 5 & 4.3$\pm$0.20 & --0.23$\pm$0.10 &  HD 38858 b&    0.0961   \\
87883 & 5015$\pm$ 5 & 4.4$\pm$0.20 &   0.00$\pm$0.10 &  HD 87883 b&   12.1      \\
95128 & 5887$\pm$ 4 & 4.3$\pm$0.20 &   0.01$\pm$0.10 &    47 Uma d&    1.64     \\
      &              &             &                &    47 Uma c&    0.54     \\
97658 & 5136$\pm$ 8 & 4.5$\pm$0.20 & --0.32$\pm$0.10 &  HD 97658 b&    0.02375  \\
128311& 4960$\pm$ 4 & 4.4$\pm$0.20 &   0.03$\pm$0.10 & HD 128311 c&    3.21     \\
145675& 5406$\pm$ 9 & 4.5$\pm$0.20 &   0.32$\pm$0.10 &    14 Her b&    4.64     \\
154345& 5503$\pm$ 6 & 4.3$\pm$0.20 & --0.21$\pm$0.10 & HD 154345 b&    1.0      \\
156668& 4850$\pm$ 5 & 4.2$\pm$0.20 & --0.07$\pm$0.10 & HD 156668 b&    0.0131   \\
186427& 5752$\pm$ 4 & 4.2$\pm$0.20 &   0.02$\pm$0.10 &  16 Cyg B b&    1.68     \\
189733& 5818$\pm$ 5 & 4.3$\pm$0.20 & --0.03$\pm$0.10 & HD 189733 b&    1.138    \\
217014& 5778$\pm$ 4 & 4.2$\pm$0.20 &   0.14$\pm$0.10 &    51 Peg b&    0.46     \\
\hline
\end{tabular}
\end{table*}

Here, we make a few comments regarding the observational data and the techniques for
determining the parameters and chemical composition. It is important
that all our determinations of both parameters and chemical composition
are performed using uniform methods.
The spectra of investigated  stars (F-G-K dwarfs) were obtained in the region
of $\lambda$ 4400--6800\,\AA\,  with typical signal-to-noise ratios S/N
of about 100-350 using the 1.93 m telescope at Observatoire de Haute-Provence
(OHP, France) equipped with the echelle-spectrograph
ELODIE \citep{baranne:96}, which gives a resolving power of R = 42000.
The initial processing of the spectra (order extraction, flat-fielding,  wavelength 
calibration and radial velocity determination) was performed with the standard online ELODIE 
reduction software, while the order deblazing and the cosmic, telluric lines and bad pixel 
removing were performed following the recipes described in \cite{katz:98}.
Further treatment of spectra (the continuous spectrum level placement,
measurement of the equivalent widths etc.) was conducted using the
{\small DECH20} software package  \citep{galazut:92}.

The atmospheric parameters of the target stars
were determined in our previous works. The methods and techniques applied are
described in
detail in \cite{mishenina:04} and \cite{mishenina:08}.
The effective temperatures \Teff\ were estimated by the calibration
of the ratio of the central depths of lines with different potentials
of the lower level developed by \cite{kovtyukh:03}, with typical internal errors better than 10 K.
The surface gravities, \logg\, for the stars
with \Teff\ higher than 5000 K were computed using two methods, namely the
iron ionization balance and the parallax. For stars with \Teff\ lower than 5000 K,
we used only the trigonometric parallax method. 
The microturbulence velocity \Vt\ was derived providing that the iron
abundance log A(Fe) obtained from Fe I lines was not correlated
with the equivalent width, EW. The adopted metallicity value
[Fe/H] was the iron abundance obtained from the Fe I lines under the  local thermodynamic
equilibrium (LTE) approximation. 
We note that the non-LTE corrections for the range of temperatures and metallicities of 
our target stars  do not exceed 0.1 dex \citep{mashonkina:11}.
In our previous papers, the external errors were estimated to be of the order 100K in \Teff\, 
0.2 dex in \logg\ and 0.1 in [Fe/H]. However, extensive comparisons with other studies 
that have a significant number of the investigated stars in common suggest that our atmospheric 
parameters were more accurate than expected. The results of these comparisons 
are shown in Table \ref{ncap}, which lists the mean differences and standard deviations. 
The mean differences reflect the biases resulting from the application of different techniques, 
atomic data and spectroscopic observations. They do not exceed 33 K in \Teff\, 0.15 in \logg\ 
and 0.05 in [Fe/H], which is indicative of the fact that our data are in good agreement 
with those reported in other studies. The standard deviations define an upper limit for 
our external errors provided that the other studies have their own (unknown) uncertainties. 
Standard deviations range from 33 to 89 K  in \Teff\, from 0.11 to 0.21 in \logg\, 
and from 0.05 to 0.09 in [Fe/H]. We expect our typical errors for a range of parameters 
to lie between these values; hence, we adopt conservative estimates 
of $\delta$\Teff $= \pm 60$ K, $\delta$\logg $= \pm0.16$ dex and $\delta$[Fe/H] = 0.06 dex 
for the accuracy of our atmospheric parameters.

\noindent
The abundances of most investigated elements were obtained
under the LTE approximation using the atmosphere models by
Kurucz (1993), the EWs of lines, and the WIDH9                     
code by Kurucz \citep{mishenina:04, mishenina:08, mishenina:12,
mishenina:13, mishenina:15a}.
The Li, O, S, Mn, and Eu abundances were
determined from the synthetic spectra calculated by a new 
version of the {\small STARSP} LTE spectral synthesis code \citep{tsymbal:96} 
with hyperfine structure
factored in for the abundances of S, Mn, and Eu. The Na, Al, Mg, and Ba
abundances were estimated under the non-LTE approximation using the
MULTI code \citep{car:86} modified by S.A. Korotin  \citep{kor:99}.
The abundance of the investigated elements were determined  by
differential analysis relative to the Sun.
The solar abundances  were calculated using the Moon and asteroid spectra, which had also been 
obtained with the ELODIE
spectrograph, and with the same atomic parameters as for the stars, both for the EW and spectral 
synthesis analysis. The differential analysis was performed for Na, Al, Mg, Mn, and Ba on 
a line-by-line basis, while for other elements averaged chemical abundances were used.

The results of the comparison of our parameters for the target stars and for the planet-hosting stars with the 
data of other authors \citep{luck:06, fuhrmann:08, gonzalez:10, dasilva:11, casagrande:11, mald:12,
ramirez:13, dasilva:15, santos:13, kang:11} are presented in Tables \ref{ncap} and A1.
There is a good agreement with our external comparison.

\begin{table}
\begin{center}
\caption[]{Comparison of our parameter determinations for all our stars and  planet-
hosting stars with the results by other authors for the $n$  common stars.}
\label{ncap}
\begin{tabular}{lrrrr}
\hline
References & $\Delta$(\Teff, K) & $\Delta$(\logg)&  $\Delta$([Fe/H]) &  n \\
\hline
{\it all our stars:}  &       &               &               &     \\
Luck \& Heiter      &33$\pm$81&0.13$\pm$0.21& 0.04$\pm$0.09&   49 \\
    (2006)       &           &               &               &      \\
Fuhrmann         &--25$\pm$33&0.08$\pm$0.17&0.01$\pm$0.06&   32 \\
    (2008)       &           &               &               &      \\
Gonzalez et al.    & --12$\pm$80&0.00$\pm$0.11&0.05$\pm$0.06&  31 \\
    (2010)       &           &               &               &      \\
Da Silva et al.    & --25$\pm$83&0.03$\pm$0.19&0.00$\pm$0.05&  41 \\
    (2011)       &           &               &               &      \\
Casagrande et al. &  --3$\pm$82&--&--&  160 \\
    (2011)       &           &               &               &      \\
Maldonado et al. & --26$\pm$89& 0.15$\pm$0.18& 0.03$\pm$0.08&  54 \\
    (2012)       &           &               &               &      \\ 
Ramirez et al.   & --14$\pm$47&0.10$\pm$0.13&  0.01$\pm$0.06&   51 \\
    (2013)       &           &               &               &      \\
Da Silva et al.    & --12$\pm$79&0.04$\pm$0.16&0.03$\pm$0.05&  74 \\
    (2015)       &           &               &               &      \\
                 &           &               &               &      \\
{\it planet host stars:} &   &               &               &     \\
Maldonado et al. &  61$\pm$55&--0.05$\pm$0.12&--0.03$\pm$0.07&  11 \\
    (2012)       &           &               &               &      \\ 
Ramirez et al.   &  18$\pm$34&--0.02$\pm$0.12&  0.00$\pm$0.06&   5 \\
    (2012)       &           &               &               &      \\
Santos et al.    &  27$\pm$65&--0.09$\pm$0.15&--0.03$\pm$0.05&  14 \\
    (2013)       &           &               &               &      \\
Kang et al.      &--32$\pm$39&--0.18$\pm$0.08&--0.05$\pm$0.05&   8 \\
    (2011)       &           &               &               &      \\
\hline
\end{tabular}
\end{center}
\end{table}

In our previous works, we analysed the errors in the abundance determinations that resulted from 
the choice of the atmospheric model parameters and EW measurements (Gaussian fitting, 
placement of the continuum) or from the synthetic spectrum fitting. For instance, 
for the n-capture element abundances \citep{mishenina:13}, we found that the
total uncertainty 
reaches 0.14--0.15 dex for the stars with low temperatures,
and 0.08--0.13 dex
for the hotter stars. However, this was done assuming larger parameter errors
than estimated here. 
Adopting errors of $\delta$\Teff $= \pm 60$ K,
$\delta$\logg\ = 0.16 dex and $\delta$[Fe/H] =  0.06 dex, we obtain new values of determination 
errors (See Tables \ref{err_7924} and \ref{err_95128}). In this case, the determination accuracy 
varies from 0.03 to 0.12 dex for all elements.

\begin{table}
\caption{Influence of the stellar parameters on the abundance determinations for an example star, 
HD 7924  (\Teff=5165 K, \logg=4.4, \Vt=1.1 km/s, [Fe/H] = --0.22).}
\label{err_7924}
\begin{tabular}{lcccc}
\hline
Species  &   $\Delta$ \Teff+60 K  & $\Delta$ \logg+0.16 & $\Delta$ \Vt+0.2 km/s & Total \\
\hline
 O {\sc i}   &  0.01  &  0.06 &   0.01 & 0.06   \\
Na {\sc i}   &  0.04  &  0.02 &   0.00 &  0.04  \\
Mg {\sc i}   &  0.04  &  0.01 &   0.02 & 0.05   \\
Al {\sc i}   &  0.04  &  0.01 &   0.02 & 0.05   \\
Si {\sc i}   &  0.01  &  0.01 &   0.03 & 0.03   \\
S {\sc i}   &  0.01  &  0.01 &   0.03 & 0.03   \\
Ca {\sc i}   &  0.05  &  0.02 &   0.07 & 0.09   \\
Mn {\sc i }  &  0.01  &  0.05 &   0.07 & 0.09   \\
Fe {\sc i }  &  0.05  &  0.01 &   0.04 & 0.06   \\
Fe {\sc ii}  &  0.05  &  0.05 &   0.04 & 0.08   \\
Ni {\sc i }  &  0.03  &  0.02 &   0.03 & 0.05   \\
Zn {\sc i}  &  0.02  &  0.03 &   0.03 & 0.06   \\
Y {\sc ii }  &  0.02  &  0.04 &   0.08 & 0.09  \\
Zr {\sc ii}  &  0.00  &  0.06 &   0.02 & 0.06   \\
Ba {\sc ii}  &  0.02  &  0.03 &   0.11 & 0.12   \\
La {\sc ii}  &  0.02  &  0.06 &   0.04 & 0.07   \\
Ce {\sc ii}  &  0.01  &  0.05 &   0.00 & 0.05   \\
Nd {\sc ii}  &  0.00  &  0.04 &   0.03 & 0.05   \\
Eu {\sc ii}  &  0.01  &  0.05 &   0.02 & 0.05   \\
 \hline
\end{tabular}
\end{table}

\begin{table}
\caption{Influence of the stellar parameters on the abundance determinations for an example star, 
HD 95128  (\Teff=5887 K, \logg=4.3, \Vt=1.1 km/s, [Fe/H] = 0.01).}
\label{err_95128}
\begin{tabular}{lcccc}
\hline
Species  &   $\Delta$ \Teff+60 K  & $\Delta$ \logg+0.16 & $\Delta$ \Vt+0.2km/s & Total \\
\hline
 O {\sc i}   &  0.01  &  0.05 &   0.01 &  0.05  \\
Na {\sc i}   &  0.04  &  0.02 &   0.00 &  0.04  \\
Mg {\sc i}   &  0.04  &  0.01 &   0.03 &  0.05  \\
Al {\sc i}   &  0.03  &  0.02 &   0.03 &  0.05  \\
Si {\sc i}   &  0.01  &  0.01 &   0.03 &  0.03  \\
S  {\sc i}   &  0.01  &  0.01 &   0.03 & 0.03   \\
Ca {\sc i}   &  0.04  &  0.02 &   0.06 &  0.07  \\
Mn {\sc i }  &  0.01  &  0.05 &   0.07 & 0.09   \\
Fe {\sc i }  &  0.03  &  0.01 &   0.04 &  0.05  \\
Fe {\sc ii}  &  0.04  &  0.04 &   0.03 &  0.06  \\
Ni {\sc i }  &  0.04  &  0.02 &   0.03 &  0.05  \\
Zn {\sc i}   &  0.02  &  0.03 &   0.03 & 0.06   \\
Y {\sc ii }  &  0.02  &  0.04 &   0.08 &  0.09  \\
Zr {\sc ii}  &  0.01  &  0.06 &   0.02 &  0.06  \\
Ba {\sc ii}  &  0.03  &  0.04 &   0.12 &  0.13  \\
La {\sc ii}  &  0.03  &  0.06 &   0.04 &  0.08  \\
Ce {\sc ii}  &  0.02  &  0.04 &   0.01 &  0.05  \\
Nd {\sc ii}  &  0.02  &  0.05 &   0.04 &  0.07  \\
Eu {\sc ii}  &  0.03  &  0.05 &   0.01 &  0.06  \\
 \hline
\end{tabular}
\end{table}

Tables A2, A3, detailing atmospheric parameters and elemental
abundances in the studied stars, are available online (see Appendix).

\section{Analysis of elemental abundance behaviour in stars with
and without planets}
\label{sec: el_analys}

\begin{figure}
\begin{tabular}{c}
\includegraphics[width=8cm]{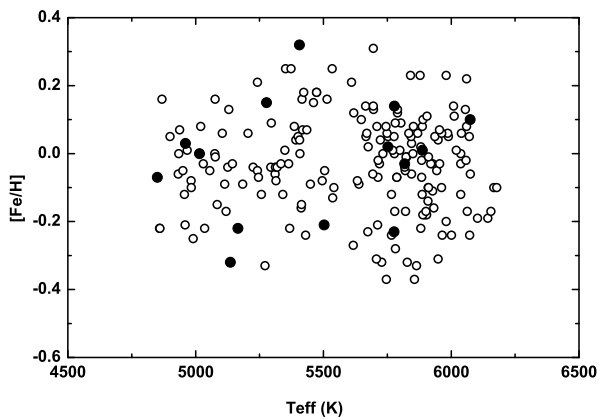}\\
\end{tabular}
\caption{Dependence of [Fe/H] on \Teff\ for our studied stars. Stars without detected planets are marked as open circles; those with planets as filled circles. }
\label{fe_teff}
\end{figure}

In Fig.  \ref{fe_teff} we present the distribution of stars with and without
planets on the [Fe/H] versus \Teff\ plane.
As can be seen from Table \ref{par_pl} and from Fig. \ref{fe_teff}, the temperatures of some 
stars with detected planets are lower than those of the solar twins selected in this paper. In addition, 
it can also be seen that several planet-hosting stars have slightly deficient metallicities. This confirms the possibility of the planetary  
formation around stars with low metallicity, as suggested by \cite{adibekyan:12b} and 
\cite{buchhave:15}, among others.

\begin{figure}
\begin{tabular}{c}
\includegraphics[width=8cm]{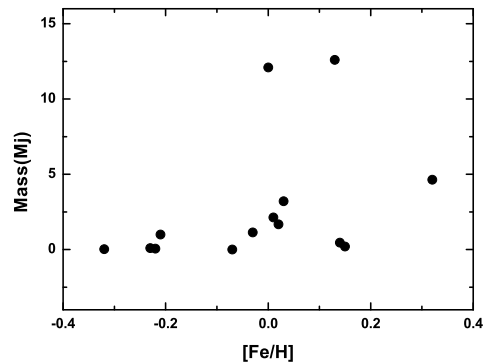}\\
\end{tabular}
\caption{Dependence of the planetary mass (in M$_J$) on [Fe/H]  for our planet-
hosting stars.}
\label{Mj_feh}
\end{figure}

A number of authors have highlighted the dependence of the mass of the
planet on the stellar metallicity \cite[e.g.][]{sousa:08, kang:11}.
The dependence of the mass of the planet (in units of Jupiter mass, M$_J$) is
shown in Fig. \ref{Mj_feh}.  We observe a negligible trend of planetary 
masses with metallicity.  In fact, the two stars with the
most massive planets (with masses of about 12 M$_J$) have a solar metallicity.
Therefore,  our small sample does not enable us to affirm the presence
of a real trend; rather, it is more likely that one does
not exist. The absence of a strong trend agrees
with the results of \cite{adibekyan:13}, who found no significant correlation between planetary 
mass and the metallicity of host stars with  Jupiter-like planets.

\subsection{Lithium}

\begin{figure}
\begin{tabular}{c}
\includegraphics[width=8cm]{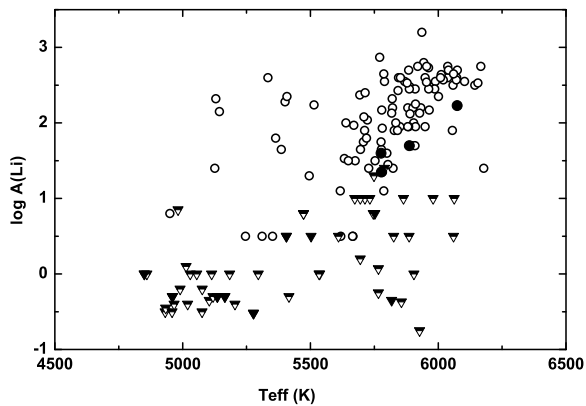}\\
\end{tabular}
\caption{Dependence of log A(Li) on \Teff\ for our studied stars.
Designations are as in Fig. \ref{fe_teff}. The stars with and without planets
having an upper limit of Li abundance marked as filled and semi-filled triangles.}
\label{li_teff}
\end{figure}

Lithium was the first element whose underabundance was
associated with the presence of planets, as reported in many studies
\citep[e.g.][]{gonzalez:00, gonzalez:08, gonzalez:10,
israelian:04,  israelian:09, delgado:14, figueira:14, delgado:15}.
The behaviour of the lithium abundance is not simple and depends on many parameters.
Lithium burns at temperatures of 2.5 x 10$^6$ K via $\alpha$-captures,
which makes it a useful tool to study mixing processes in stars.
Apart from convective overshooting, there are also other mechanisms that result 
in changes of the lithium abundance, such as meridional circulation
or diffusion. Lithium appears to be the most thoroughly investigated element,
a large number of studies have focused on the relationship
between the lithium abundance in a star and its properties, such as mass, age, 
rotation, chromospheric activity, etc. In this paper, we compare 
the lithium behaviour in planet-hosting and other dwarfs from 
our group of target stars  (Fig. \ref{li_teff}).

The lithium abundances were determined by synthetic spectra tools using
a new version of the {\small STARSP} LTE spectral synthesis code \citep{tsymbal:96}
and the list of lines in the region of the Li I line 6707 \AA\ from
\cite{mishenina:97}. In this study, the lithium abundances of 84 stars were derived  
for the first time, while those of the remaining 56 stars were taken from our 
previous studies  \citep{mishenina:08, mishenina:12}.
In spite of the errors in the lithium abundance estimate, arising for various reasons, 
we emphasize here that the effective
temperature is the key parameter, whose accuracy is crucial
in the lithium abundance determination. Recently, \cite{figueira:14}
applied multivariable regression to study the role of presence of planets on 
the lithium abundance in stars. These authors, by simultaneously considering 
the impact of different parameters on lithium abundances, concluded that planet- 
hosting stars display a depletion in lithium.

As we can be seen from Table \ref{li_comp}, the difference between
the values obtained by different authors is within 0.1 dex. We have not determined 
the lithium abundance for HD 87883 because the spectrum is distorted in the region of 
the lithium lines.

\begin{table}
\caption{Stellar age and lithium abundance in the stars with detected planets. }
\label{li_comp}
\begin{tabular}{rcrrrrl}
\hline
 HD     &Li  & Li$_{up}$& 4    &5    & 6  & Age (Gyr) \\
\hline
 3651   &--  &--0.52& 0.39 &--   &     &  7.89:\\
 7924   &--  &--0.30& --   &--   &     &  1.03:\\
 9826   &2.23&    --& 2.28 &--   & 2.47&  5.31\\
 38858  &1.60&    --&    --&1.49 & 1.64&  4.03:\\
 87883  &    &    --&    --&  -- &     &  8.92\\
 95128  &1.70&    --&  1.72&  -- &     &  9.35:\\
 97658  &--  &--0.30&--    &--   &     &  0.27:\\
 128311 &--  &--0.30&   -- &  -- & $<$--0.37&  6.03  \\
 145675 &--  &  0.50&--    &--   & $<$--0.02&  4.26: \\
 154345 &--  &  0.50&  --  &--   &          &  - \\
 156668 &--  &  0.00&    --&   --&          &  0.32: \\
 186427 &--  &  0.80&  0.75&   --& $<$0.60  &  6.85 \\
 189733 &--  &--0.35& --   &   --&          &  - \\
 217014 &1.35&    --&  1.21& 1.29& 1.30     &  8.09 \\
\hline
\end{tabular}
\\
Notes: Li-- our lithium abundance determination; \\
 Li$_{up}$ -- upper limit of the lithium abundance; \\
4 --\cite{ramirez:12}, 5 -- \cite{delgado:15},  \\
6 -- \cite{israelian:04}, 7 -- stellar age,  \\
 the values of age marked with (:)  correspond to more uncertainties
in age determinations. \\
\end{table}

\begin{figure}
\begin{tabular}{c}
\includegraphics[width=8cm]{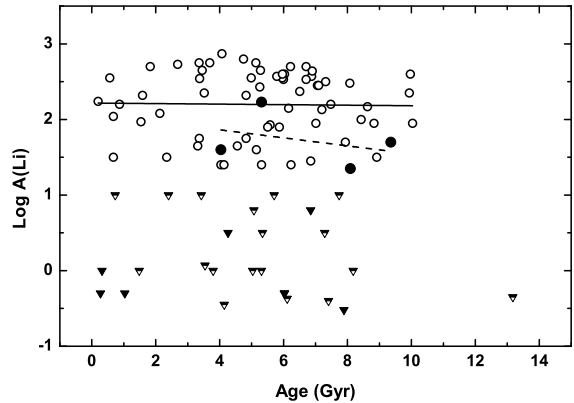}\\
\end{tabular}
\caption{Relation between the lithium abundance log A(Li) and age for our studied stars.
Designations are as in Fig. \ref{li_teff}. }
\label{li_age}
\end{figure}

A comparison of the lithium patterns in stars with and without
detected planets is presented in Fig. \ref{li_teff}.  As can be seen
from the figure, the lithium abundances of planet-hosting stars are lower 
than those in the dwarfs without detected planets.
In our example, the lithium abundance of planet-hosting stars is not greater 
than 1.7, with the only exception being HD 9826 (\Teff\ = 6074 K), and for most 
of these stars we found only the upper limit of lithium abundance 
(it should be noted that these values depend on the signal-to-noise ratio and 
resolution of the spectra).  
There are two possible explanations for our results (i.e. the low lithium abundance).
First, we had a small selection of planet-hosting stars, namely 14. 
Secondly, the temperature of several planet-hosting stars 
was lower than \Teff\ $<$ 5500 K, the temperature at which convection is the main mechanism for
the lithium depletion. However, the lithium abundances in hot planet-hosting 
stars are lower than in dwarfs with similar values of \Teff. These results are in good 
agreement with those obtained in earlier studies \cite[e.g.][]{israelian:04}.
These authors suggested that there is a likely excess lithium depletion in planet-hosting 
stars with \Teff\ in the range 5600--5850 K, but no significant differences in
the stars with \Teff\ in the range 5850--6350 K. In a recent study \citep{figueira:14},
 it was reported that such an offset is negative indicating an enhanced depletion in 
planet-hosting stars; it was also shown that the results obtained were statistically 
meaningful. For a larger sample of 326 Main-Sequence stars with and without 
planets and a \Teff\ range of 5600--5900 K \citep{delgado:14}, the lithium abundance
determinations suggested that the amount of lithium depletion in planet-hosting 
solar-type stars is higher when the masses of planets exceed 1M$_J$. Finally, 
\cite{delgado:15} found that the lithium abundances in the stars hosting hot Jupiters
 with a \Teff\ range of 5900--6300 K is by 0.14 dex lower than in stars without
detected planets. In our sample, there are four stars with planetary systems in the 
temperature range 5700--6100 K with masses greater than 1M$_J$. There is only one 
star, namely HD9826 with \Teff\ = 6074 K and a planetary mass of about 12 M$_J$ 
that exhibits a high lithium abundance of 2.23 dex.
A different view of the lithium behaviour has, however, been expressed in a number of 
papers \cite[e.g.][etc]{baumann:10, carlos:16}. 
In particular, some authors found strong evidence that lithium depletion increases 
with age, a finding that is pertinent to stellar evolution rather than to the presence of planets.  
Fig. \ref{li_age} shows the dependence of lithium abundance on stellar age for 
our sample. The age determinations were taken from \citep{mishenina:13}. The reliable 
age determinations were made using tracks \cite{mowlavi:12} and only for the objects 
fitted by the correct isochrones set with pair-wise Euclidean distances in two-dimensional 
space (presented by $T_{\rm eff}$/$T_{\odot}$ and $L/L_{\odot}$) smaller than 0.02. 
The age of the stars whose determinations were made with increments smaller than 
0.05 are marked with a colon (:) in Table \ref{li_comp}.

As we can be seen from Fig. \ref{li_age}, 
there is no dependence of lithium on the age of the stars with detected lithium abundance. 
The slope for the stars without 
planets is --0.003 $\pm$0.023 and it is --0.053 $\pm$0.100 for the stars with planets. 
Unfortunately, we have only four stars with planets and detected lithium abundance. 
Two of the stars, HD 9826 and HD 217014, with detected lithium (2.23 and 1.35) have ages 5.31 Gyr 
and 8.09 Gyr, respectively. For the other two stars, the age determinations are uncertain. 
Hence, it is not possible to draw any reliable conclusions about the correlation between the 
lithium abundance and age. As we consider such a small number of stars for which lithium is detected, 
our results sustain the hypothesis regarding lithium dilution in stars with planets 
to some extent, while the existence of a correlation between the lithium abundance and 
age is supported to an even lesser extent.

\begin{figure*}
\begin{tabular}{cc}
\includegraphics[width=5.5cm]{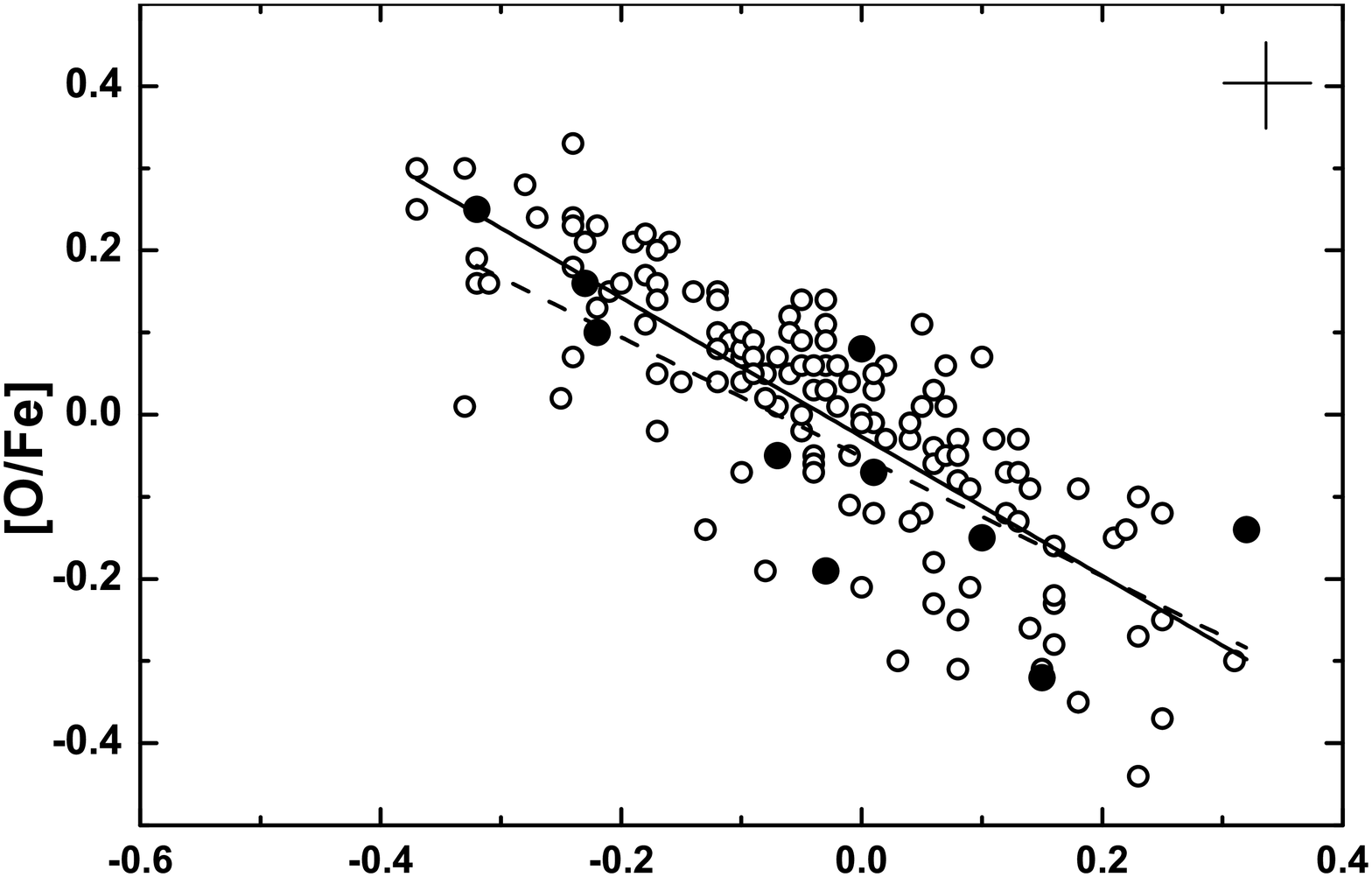} &\includegraphics[width=5.5cm]{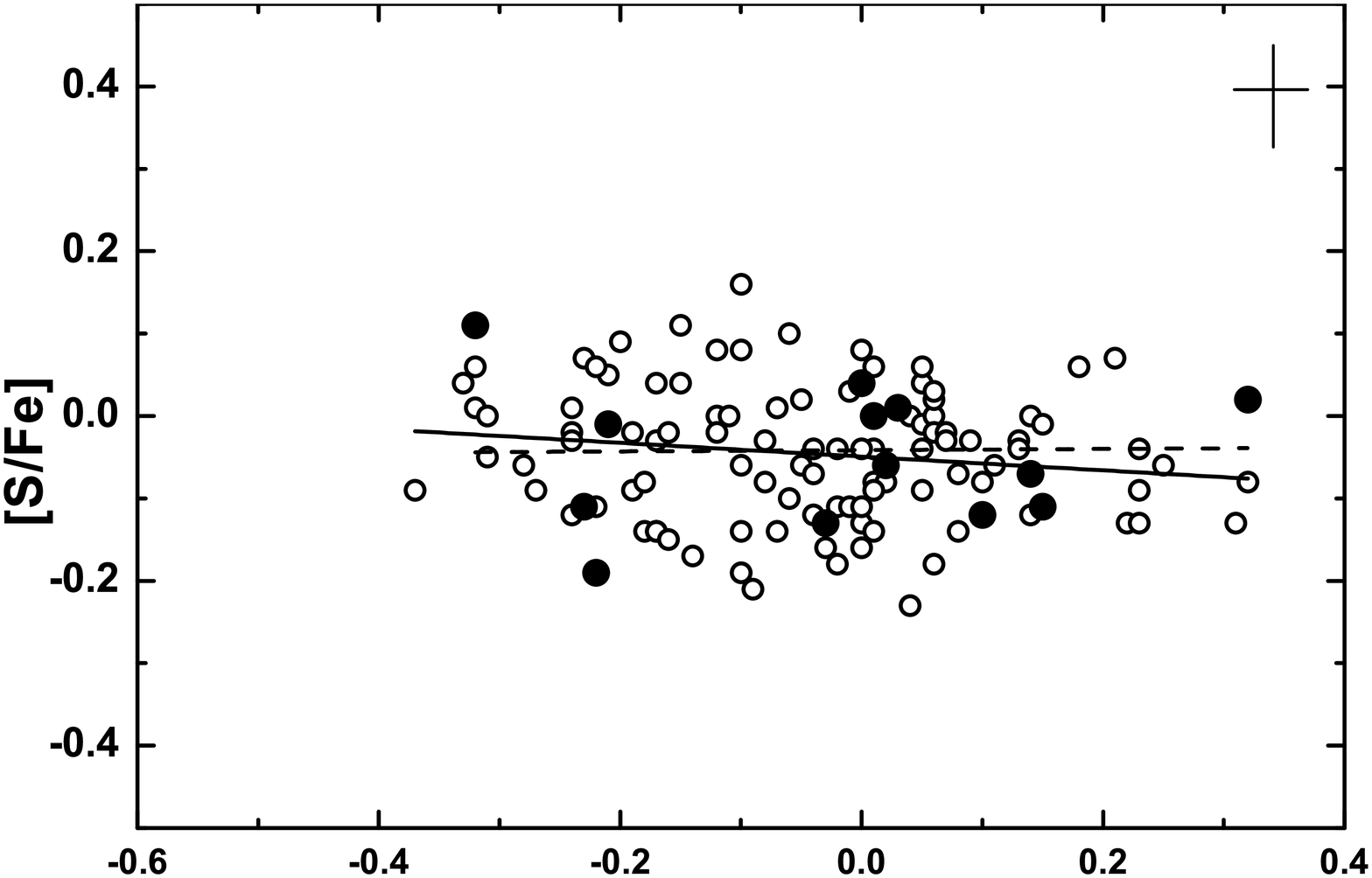} \\
\includegraphics[width=5.5cm]{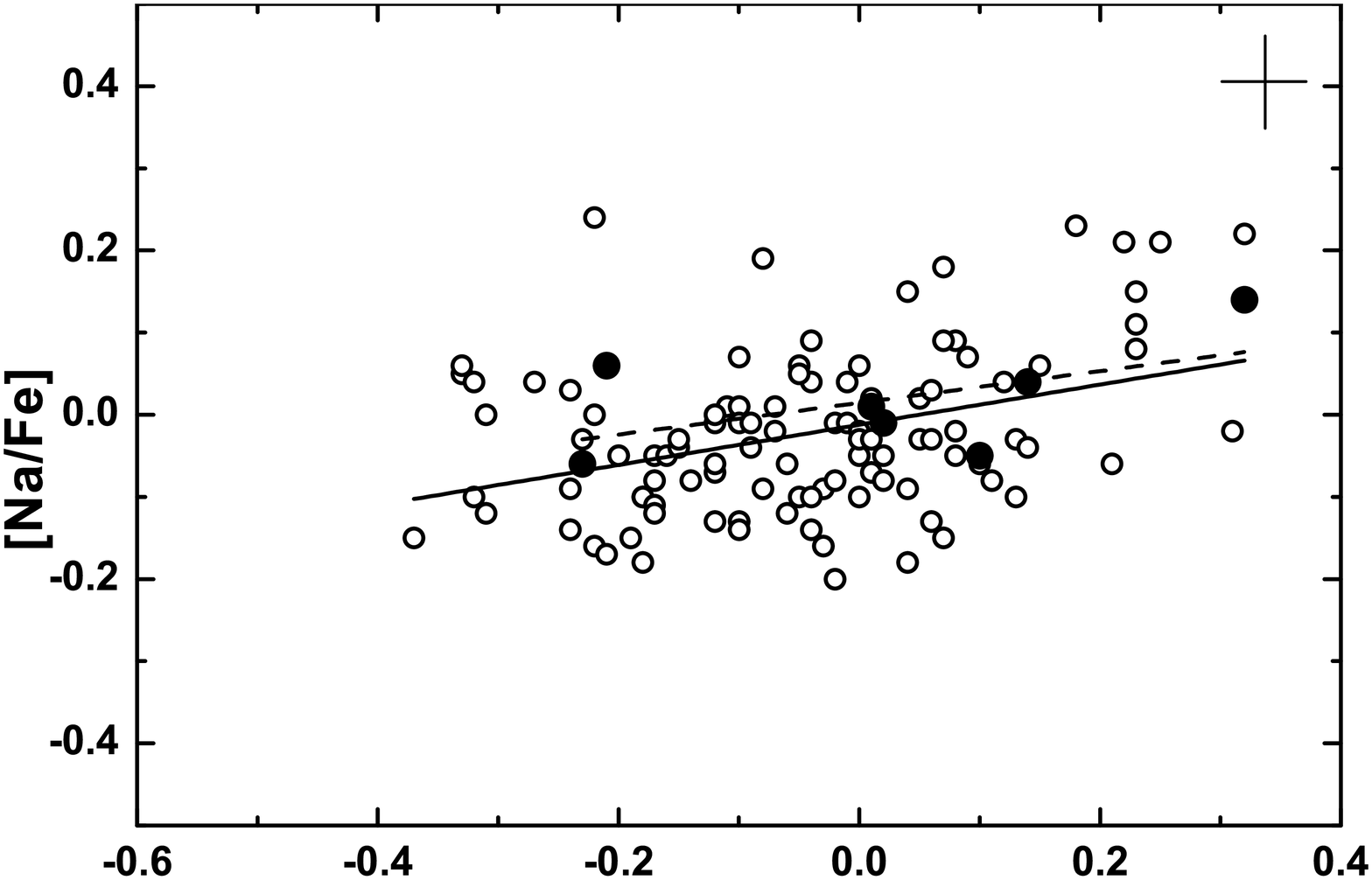} &\includegraphics[width=5.5cm]{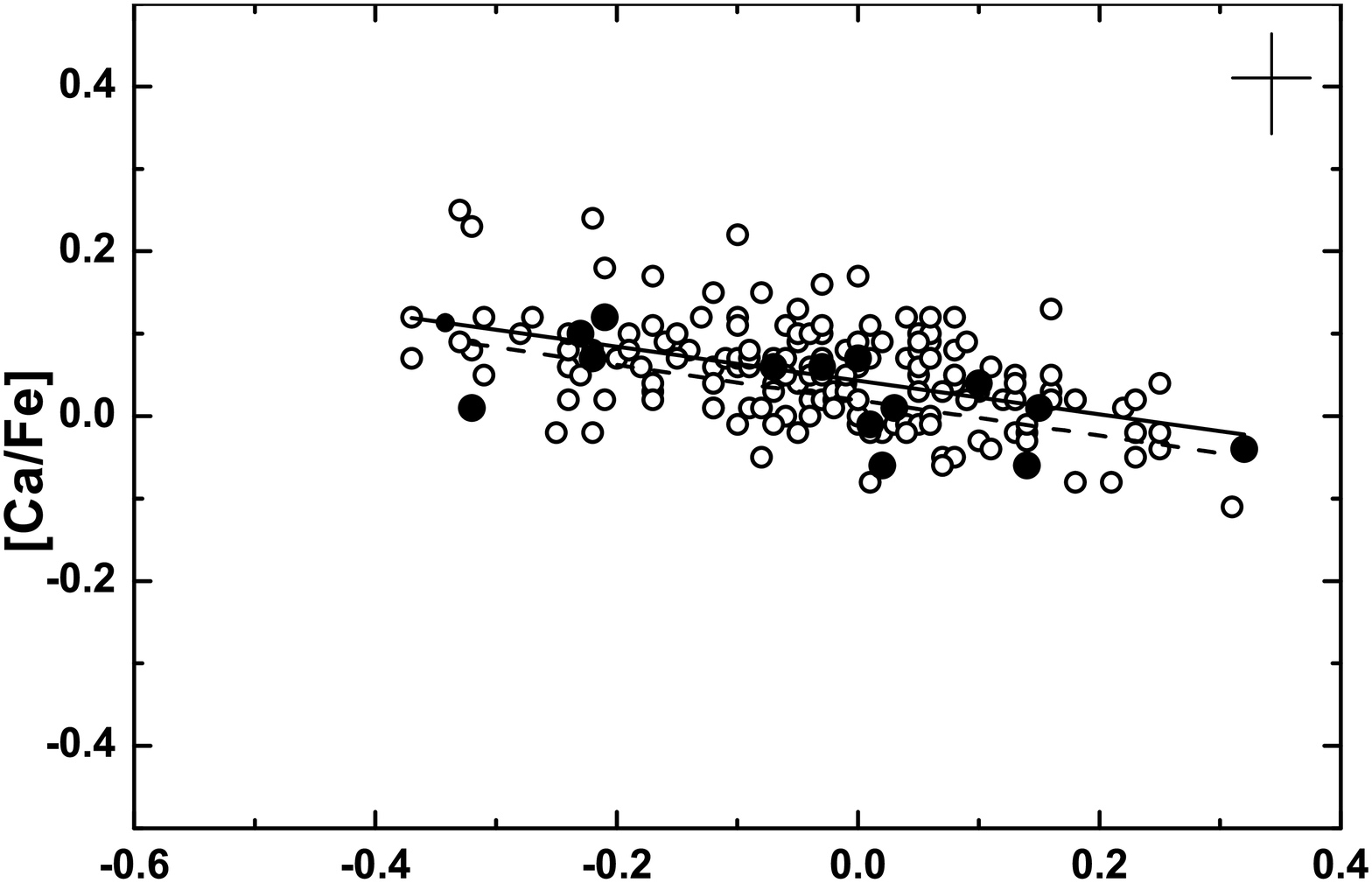} \\
\includegraphics[width=5.5cm]{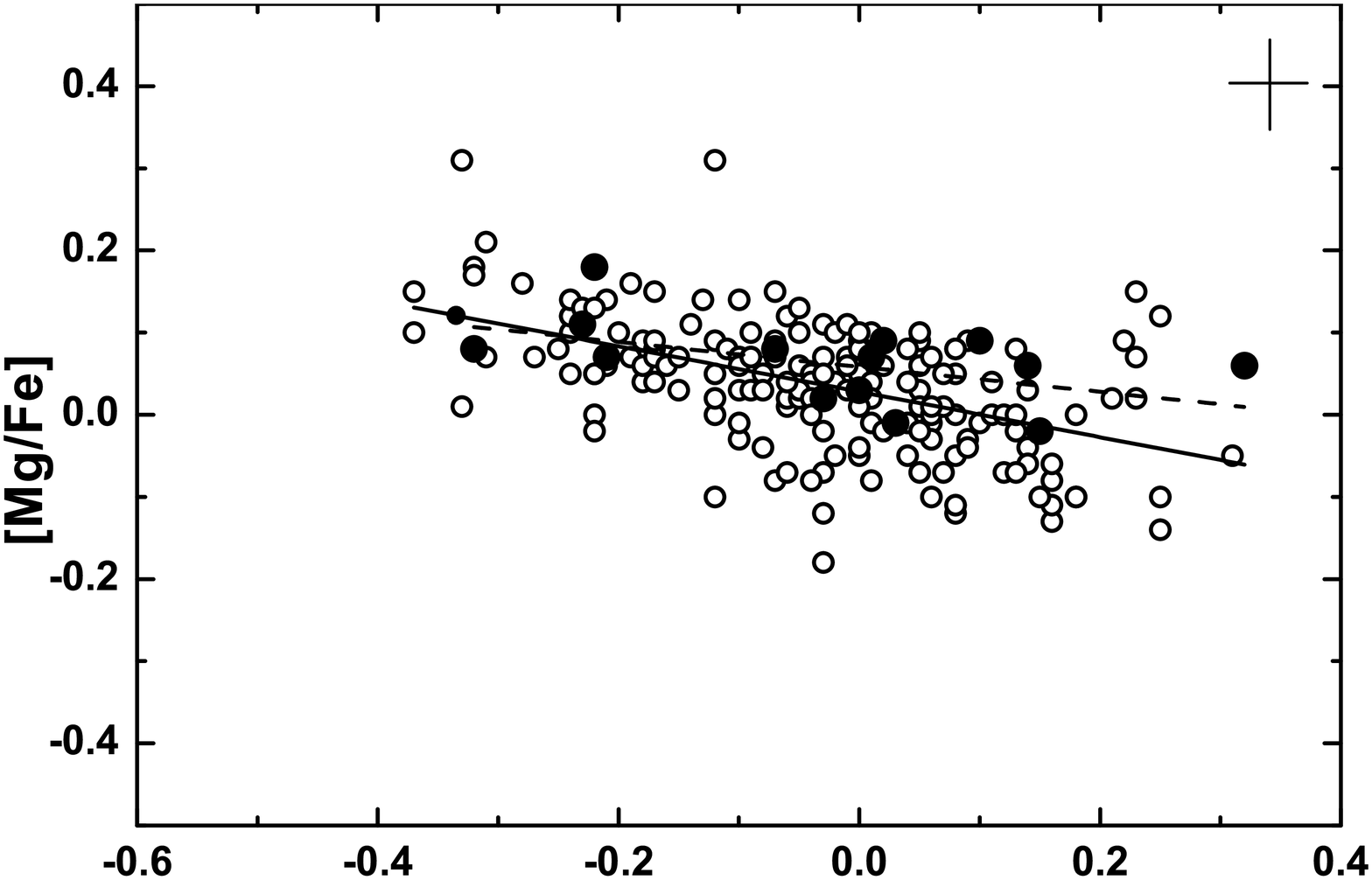} &\includegraphics[width=5.5cm]{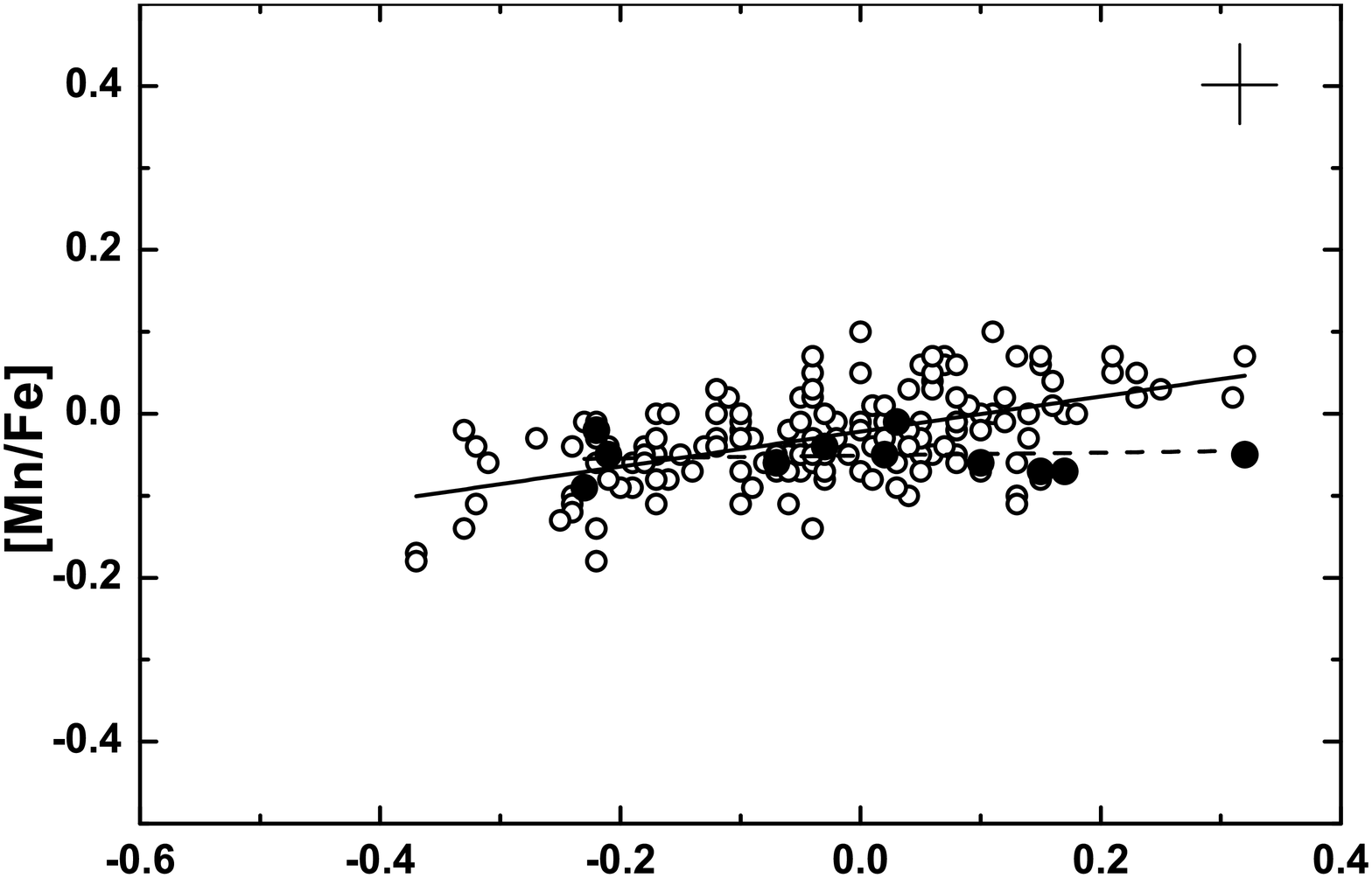}  \\
\includegraphics[width=5.5cm]{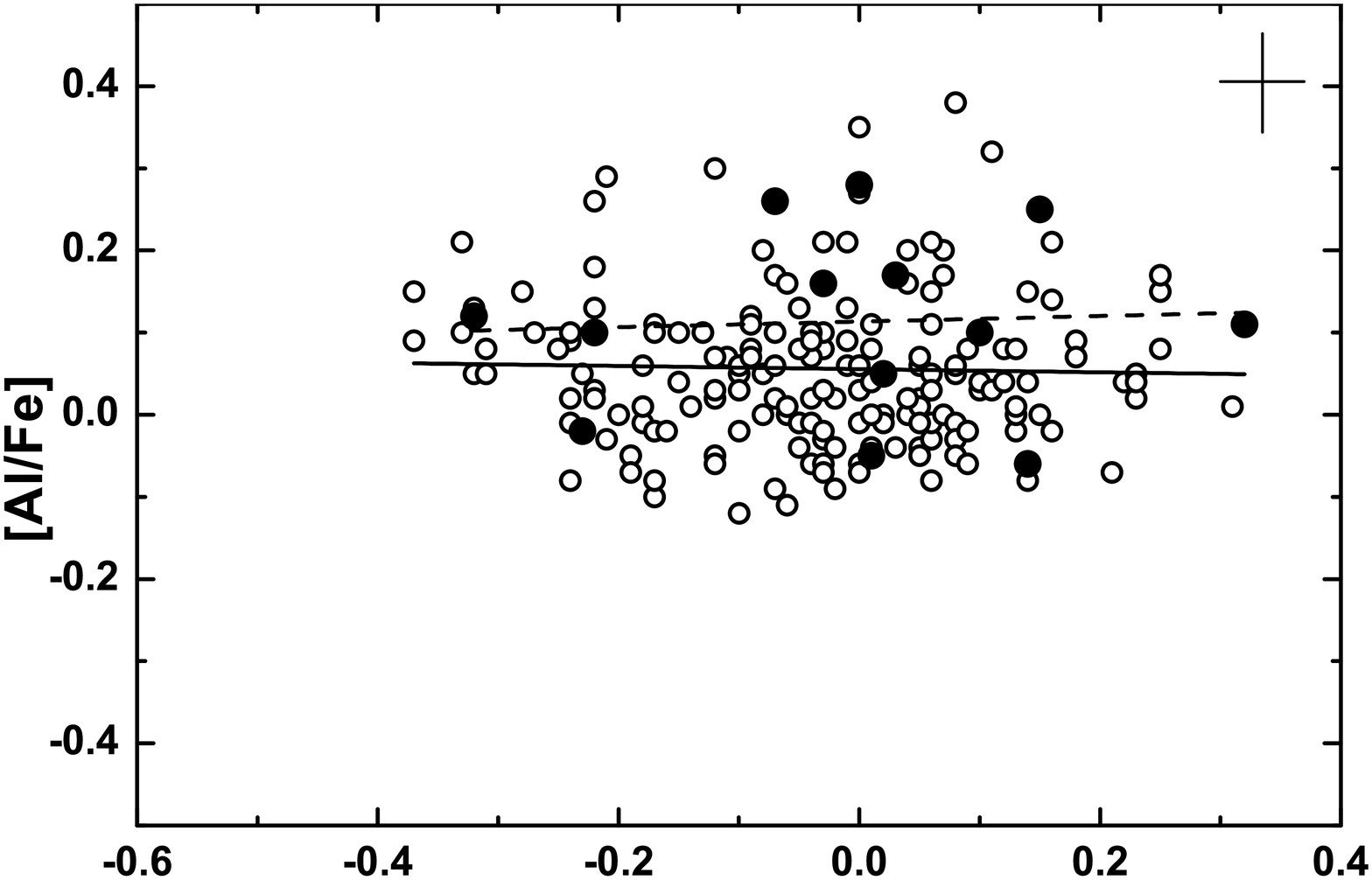} &\includegraphics[width=5.5cm]{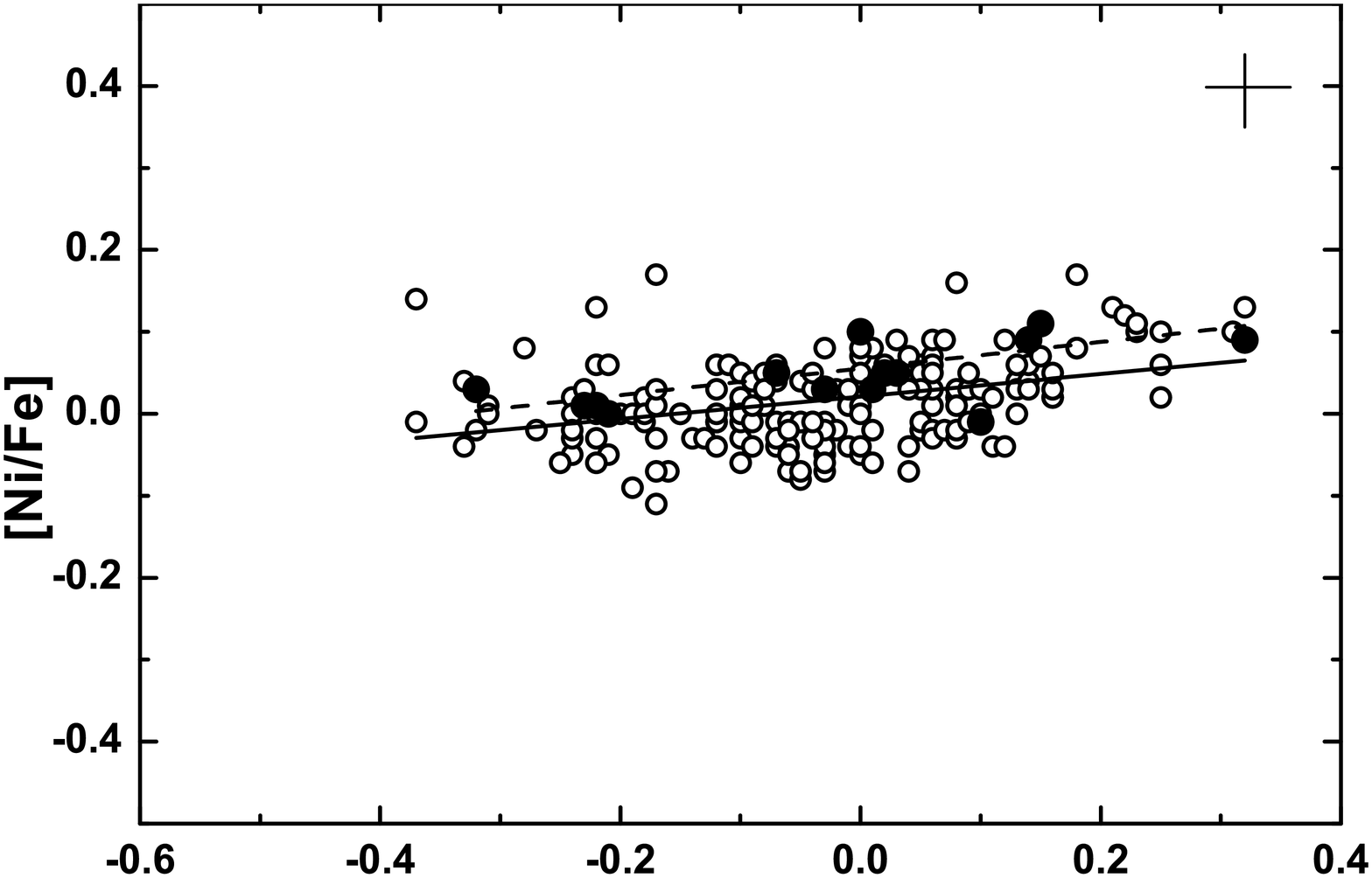}\\
\includegraphics[width=5.5cm]{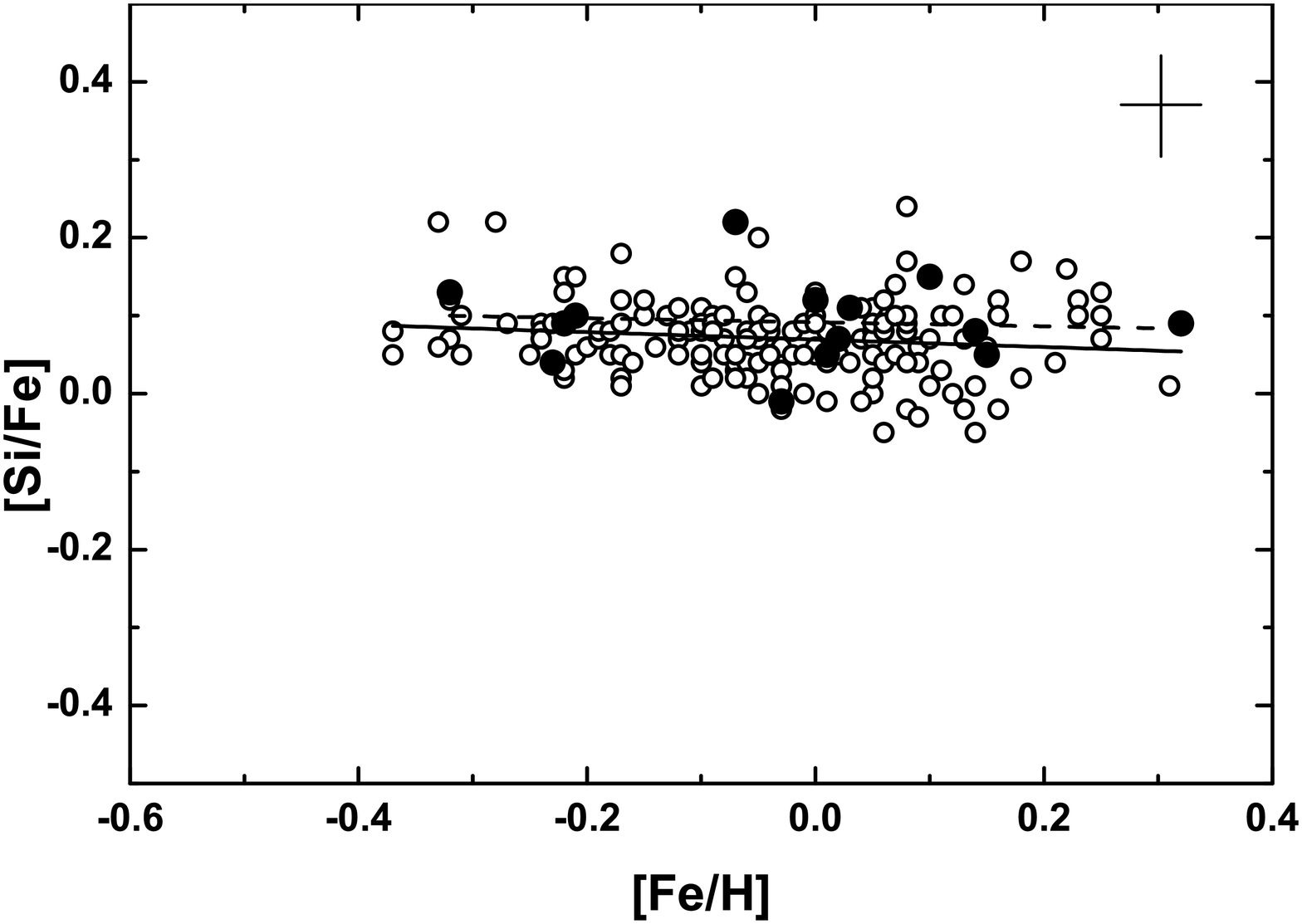} &\includegraphics[width=5.5cm]{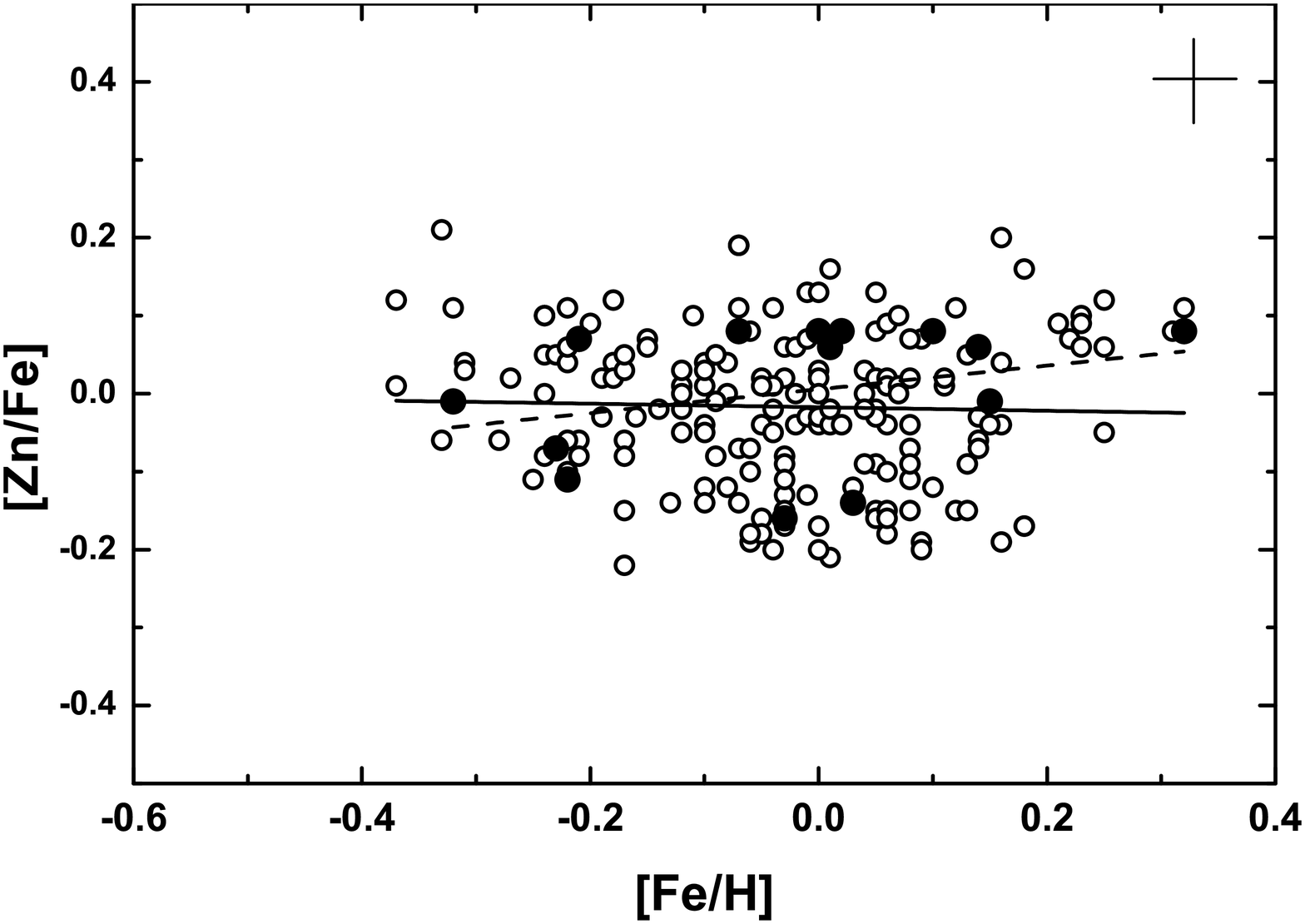}\\
\end{tabular}
\caption{Abundance [El/Fe] trend  versus metallicity [Fe/H].}
\label{abun1}
\end{figure*}

\begin{figure*}
\begin{tabular}{cc}
\includegraphics[width=6.0cm]{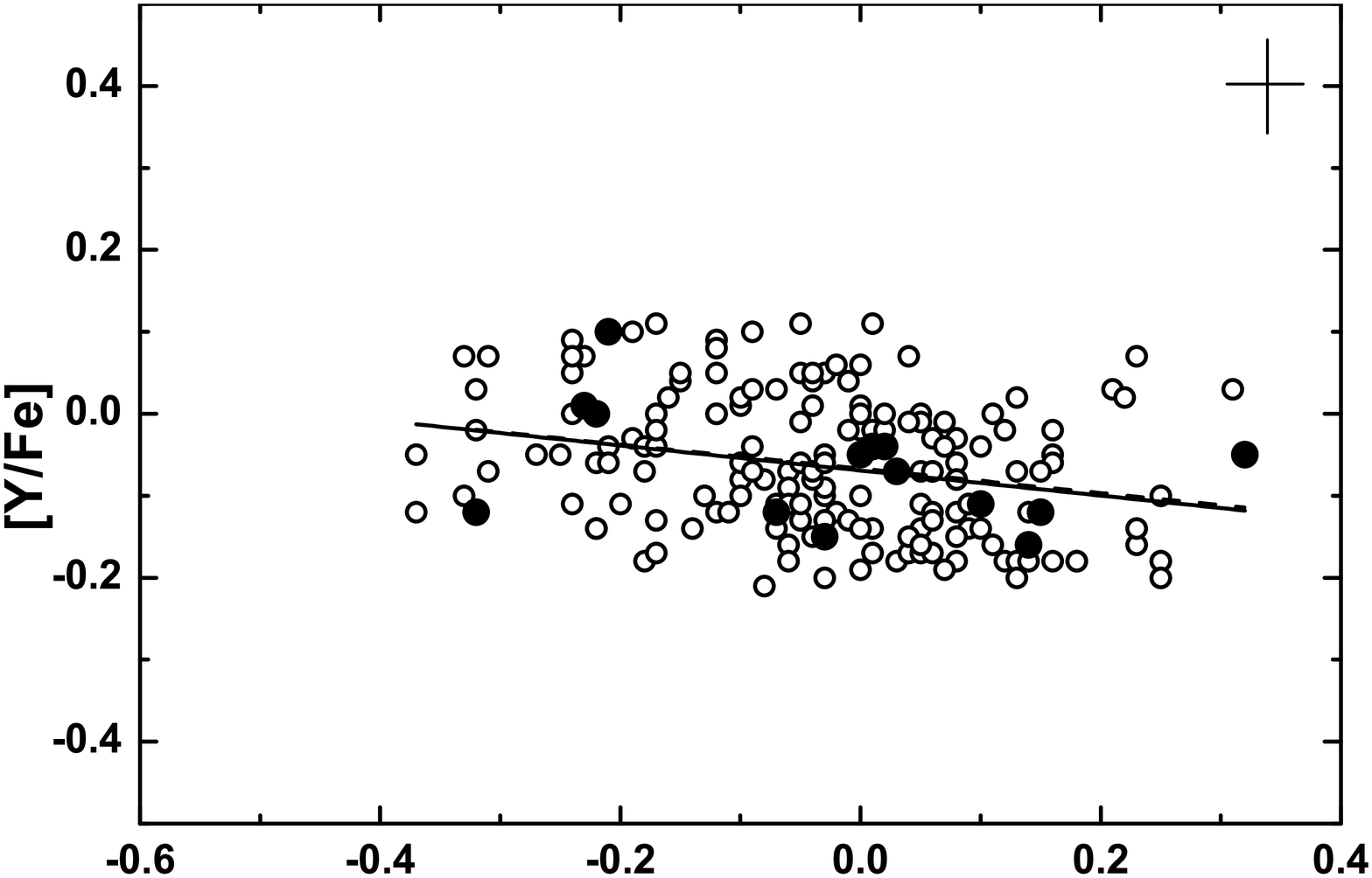} &\includegraphics[width=6.0cm]{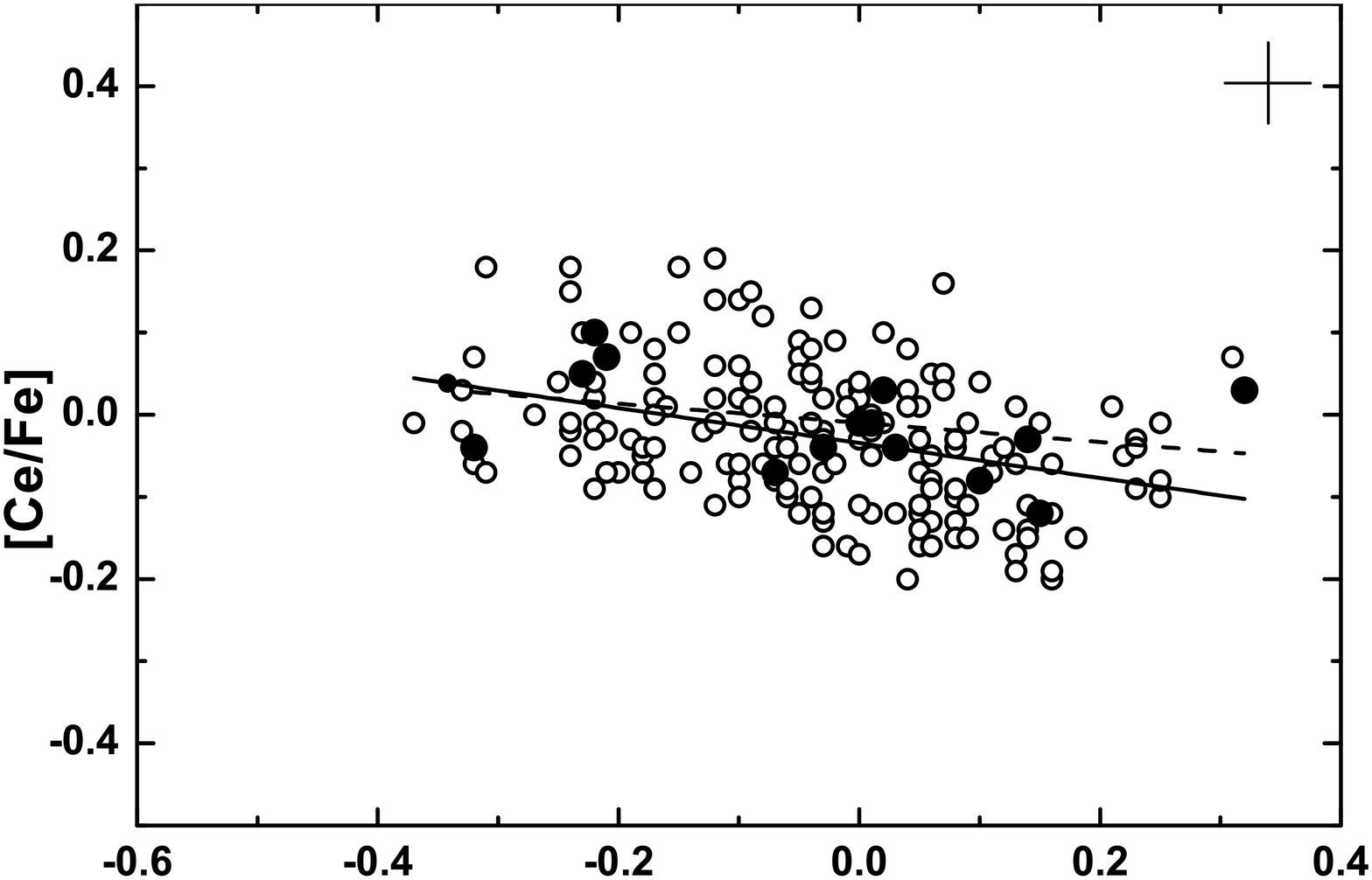}\\
\includegraphics[width=6.0cm]{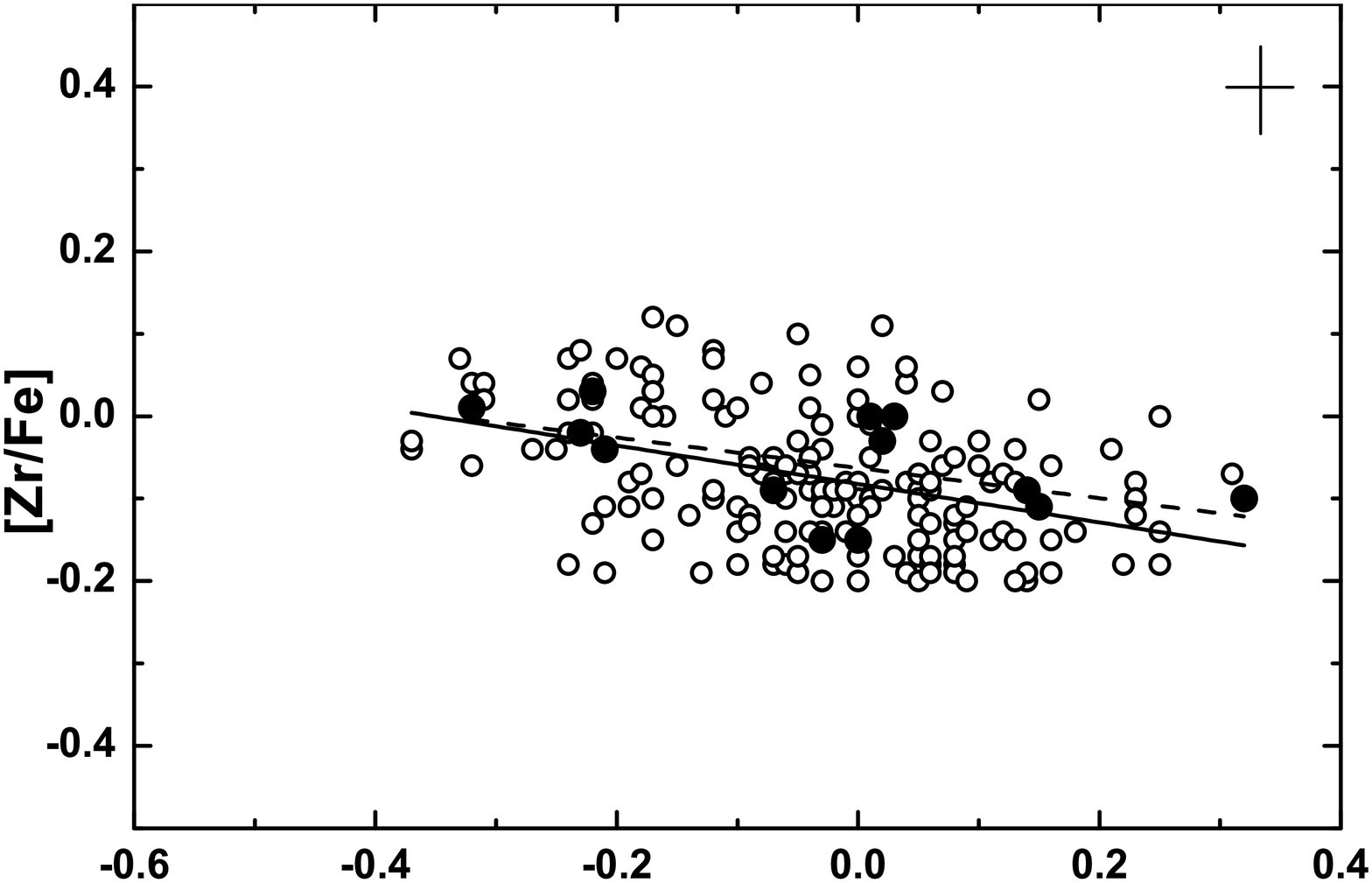} &\includegraphics[width=6.0cm]{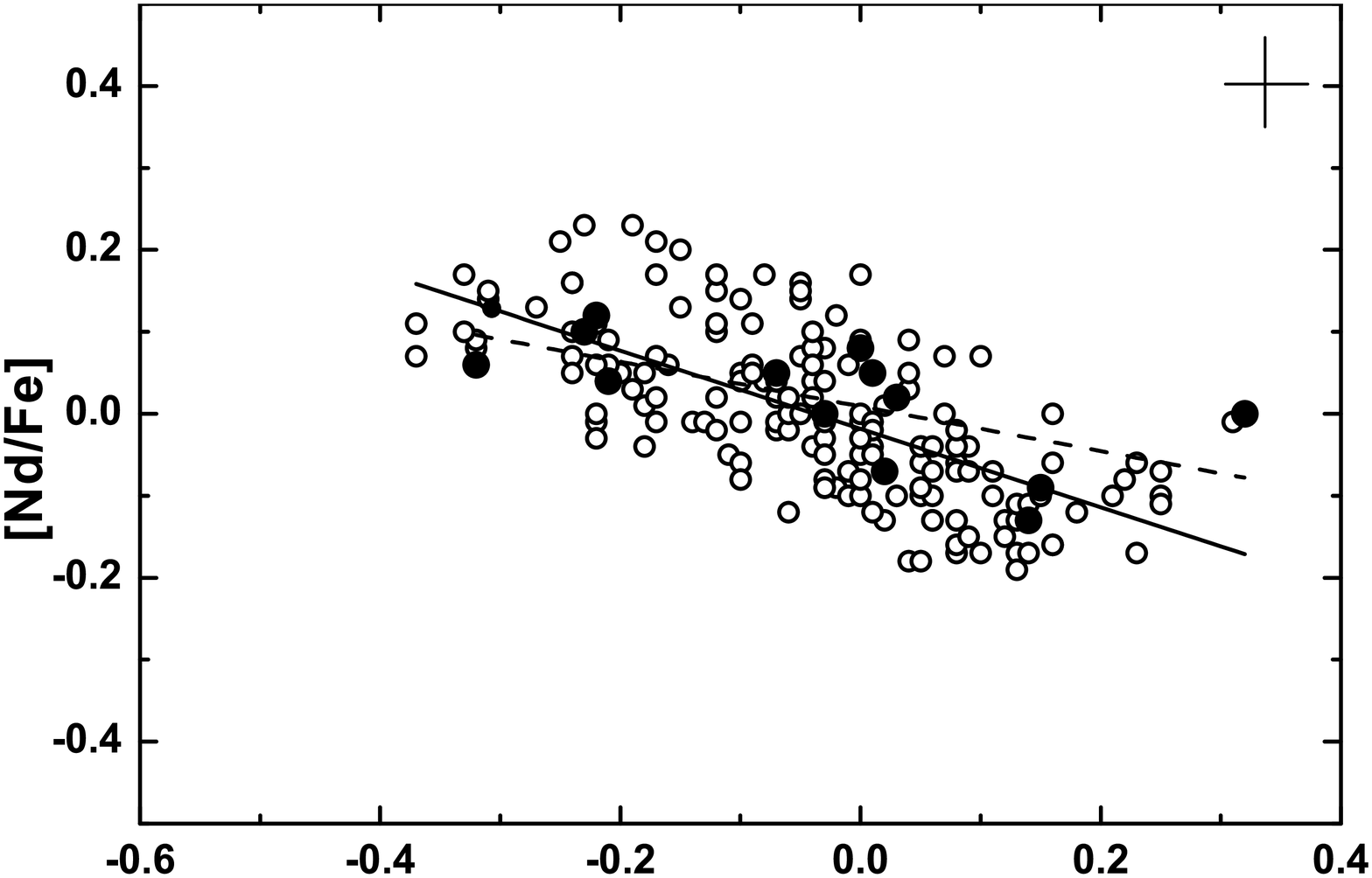}\\
\includegraphics[width=6.0cm]{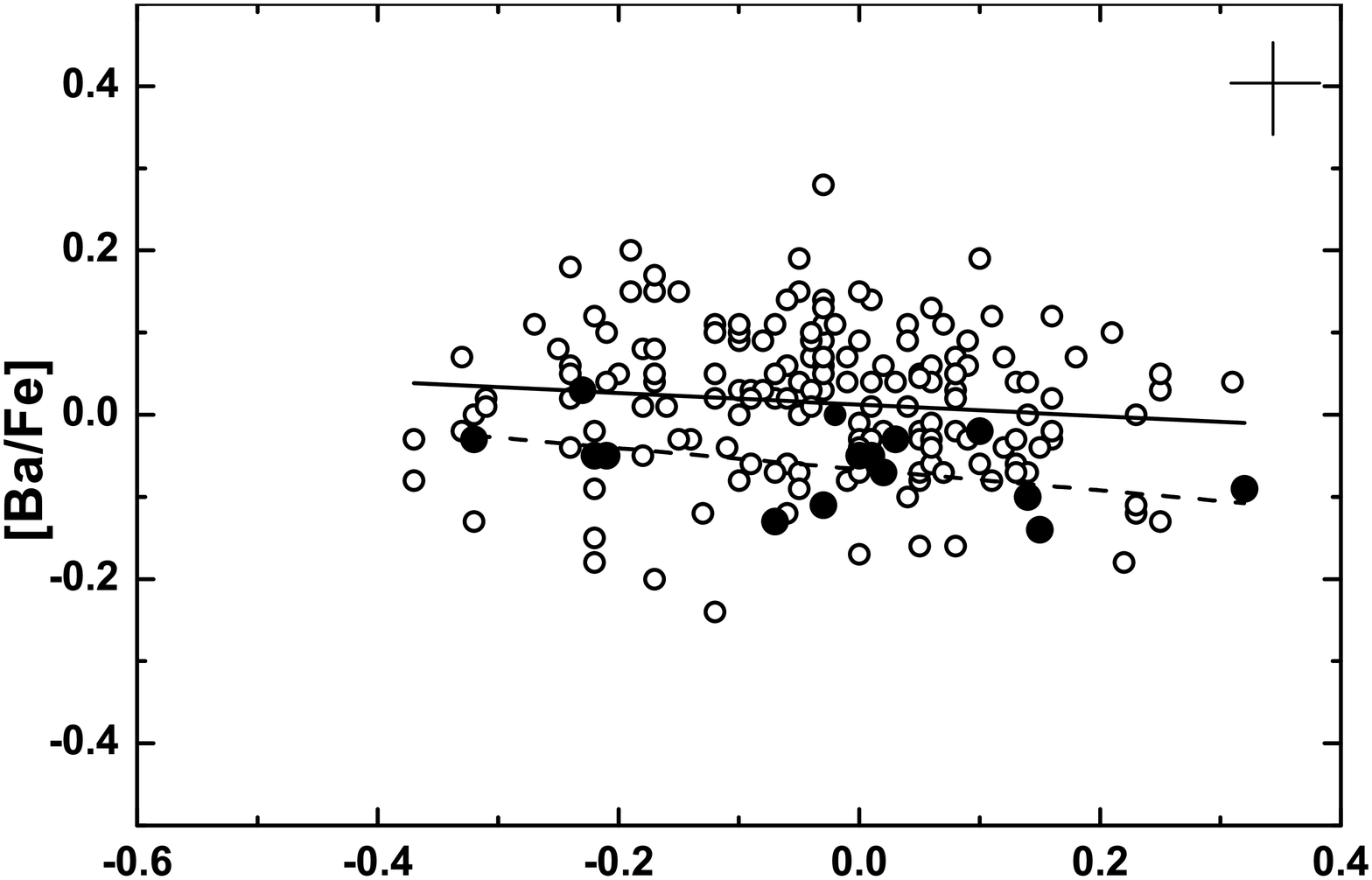} &\includegraphics[width=6.0cm]{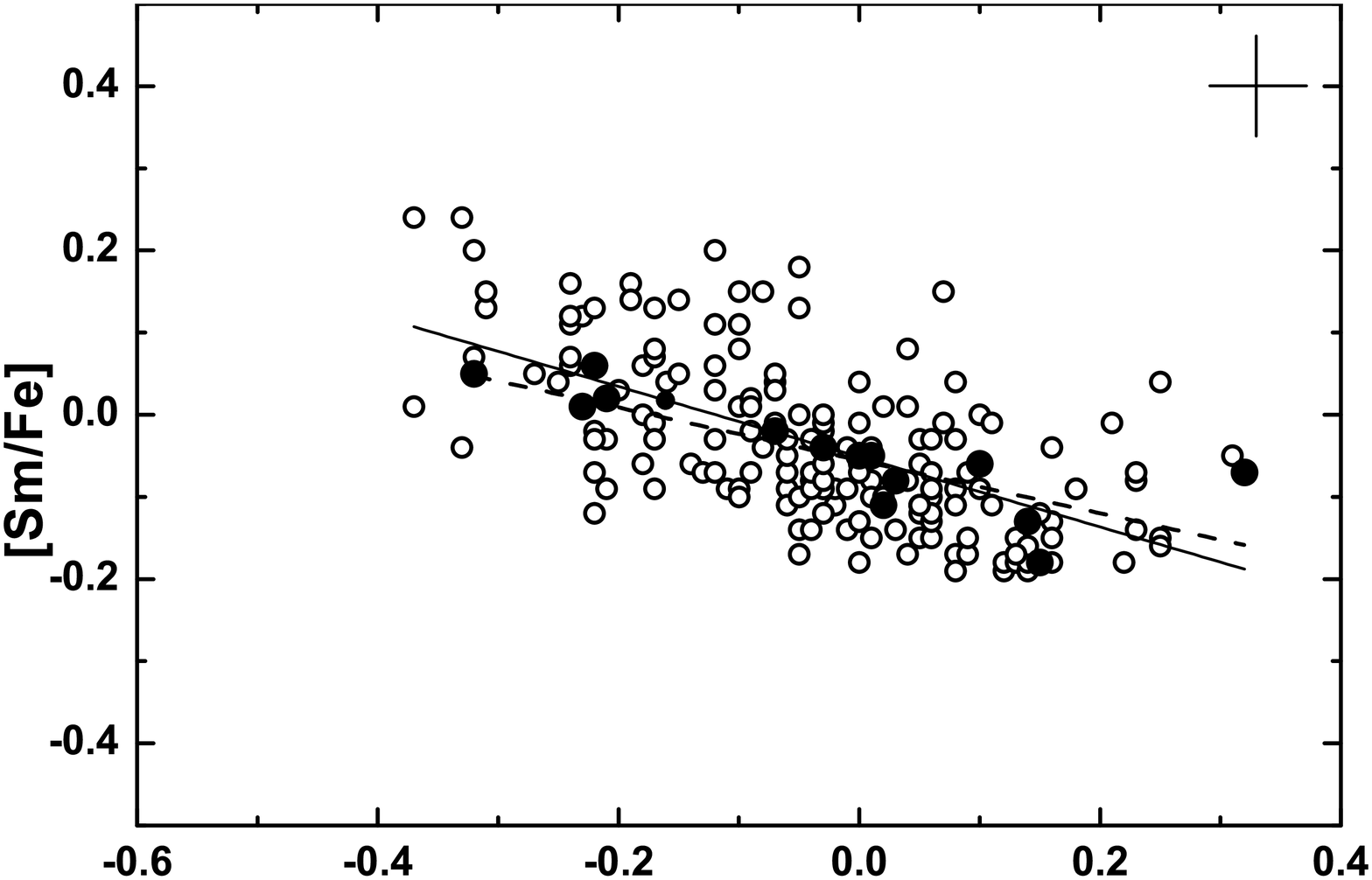}\\
\includegraphics[width=6.0cm]{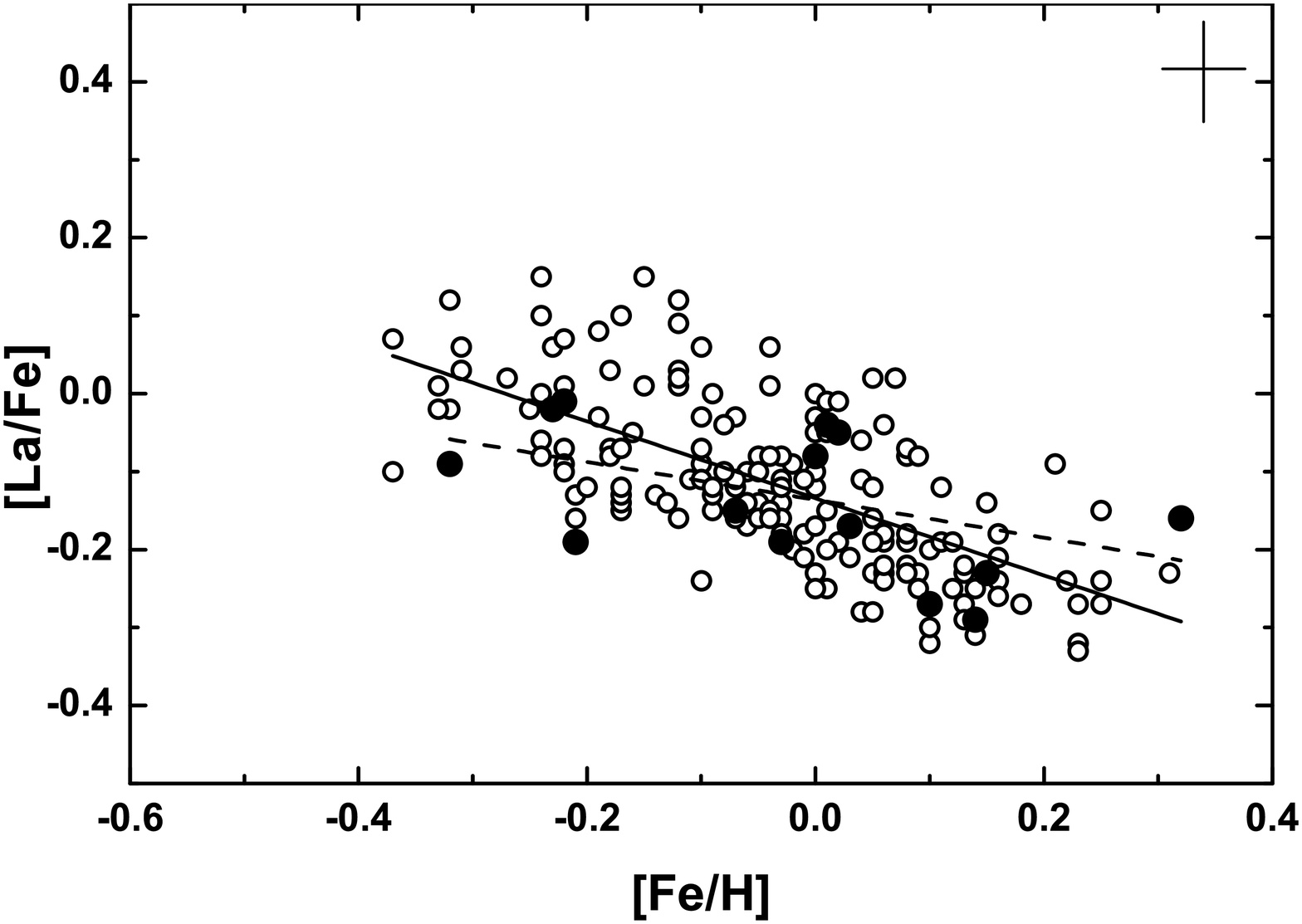} &\includegraphics[width=6.0cm]{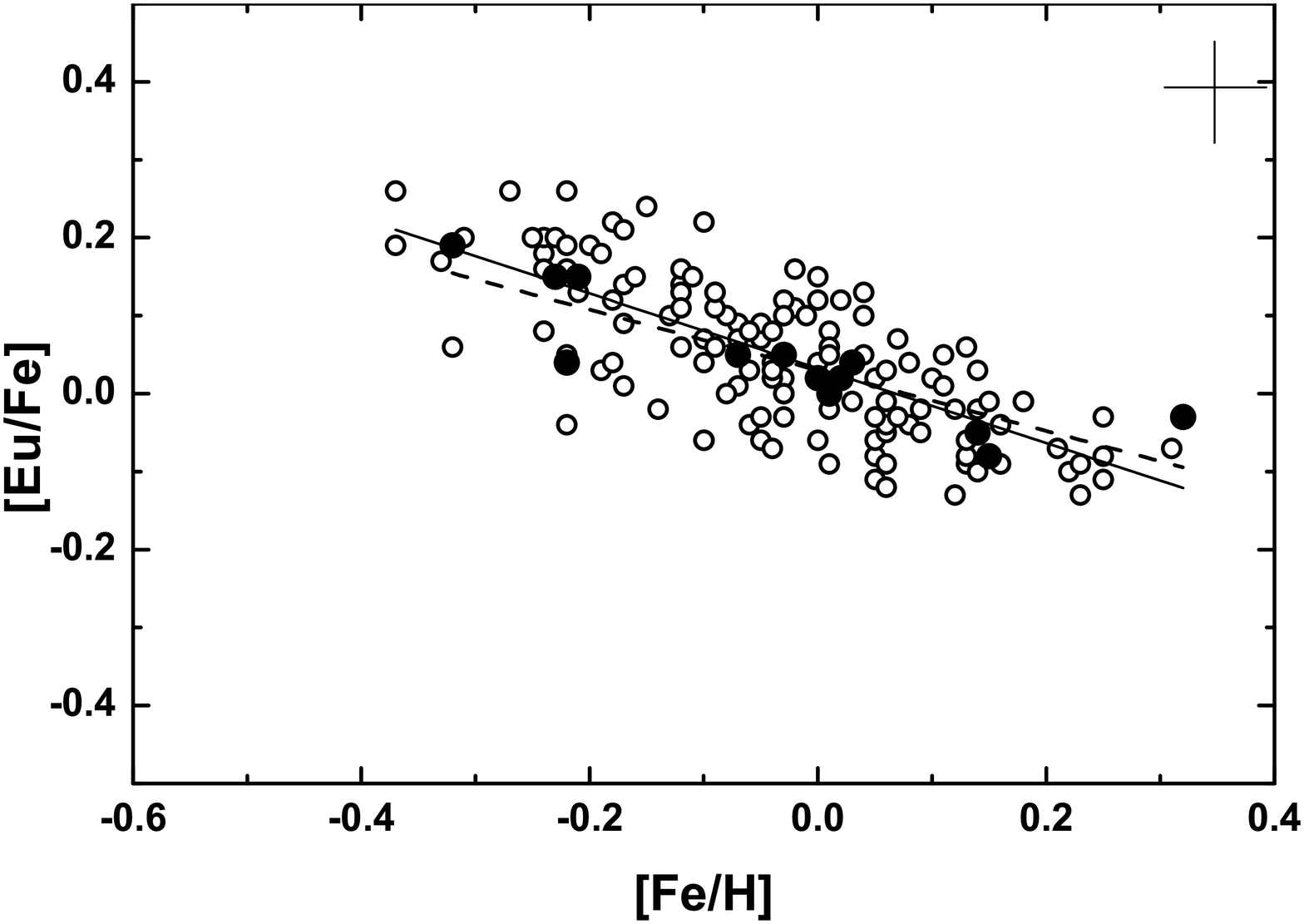}\\
\end{tabular}
\caption{Abundance [El/Fe] trend  versus metallicity [Fe/H].}
\label{abun2}
\end{figure*}

\subsection{$\alpha$-elements (Mg, Si, Ca) }

In Fig. \ref{abun1} and \ref{abun2}, we present the dependence of the [El/Fe] abundance 
ratios on the stellar metallicity. 
As can be seen from these figures, there are no
significant differences between stars with and without planets.
It is likely that we cannot see the same enhancement in $\alpha$-elements as 
it was shown in \citep{adibekyan:12a, adibekyan:12b} owing to the lack of low-metallicity stars 
in our target sample.

\begin{figure}
\begin{tabular}{c}
\includegraphics[width=8cm]{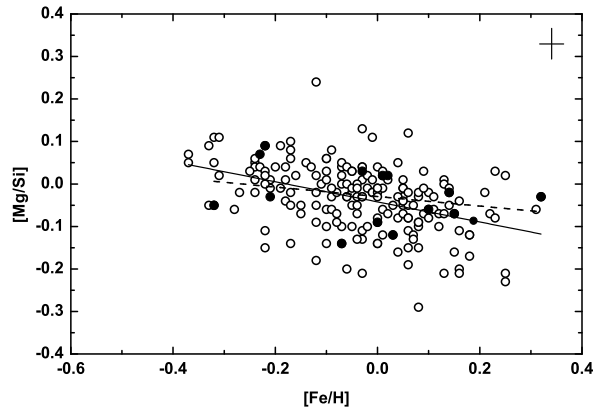}\\
\end{tabular}
\caption{Dependence of [Mg/Si] on [Fe/H] for our target stars.
Designations are as in Fig. \ref{fe_teff}.}
\label{mgsife}
\end{figure}

\begin{figure}
\begin{tabular}{c}
\includegraphics[width=8cm]{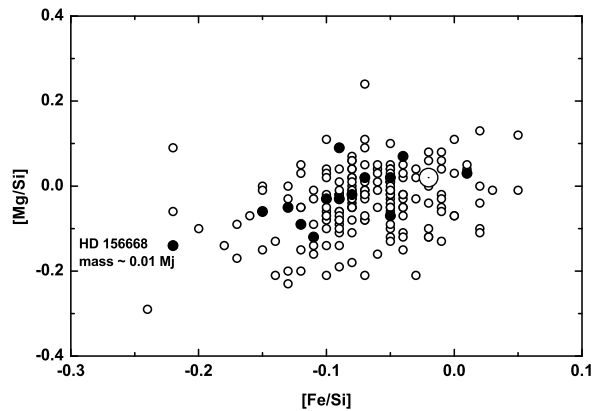}\\
\end{tabular}
\caption{Dependence of [Mg/Si] on [Fe/Si] for our target stars. Designations are as in Fig. \ref{fe_teff}. The solar value is marked as a circle with a point.}
\label{mgsi}
\end{figure}

If the mass and radius are known, it is possible to derive constraints on the
interior structure of exoplanets \citep{dorn:15}. Using an inversion
method based on Bayesian analysis, 
these authors investigated constrains on the interior structure of
terrestrial exoplanets, 
in the form of the chemical composition of the mantle and the core size.
In particular, 
they concluded that stellar elemental abundances (Fe, Si, Mg)
are the principal constraints for 
reduction of  degeneracy in interior structure models and limiting factors
for the mantle composition. Therefore, the study of these elements is
essential in order to detect either small-mass rocky planets or rocky cores of
massive planets based on the stellar chemical composition.
\cite{santos:15} derived stellar parameters and chemical
abundances for Fe, Si, Mg, O, and C in three stars
hosting low-mass planets, rocky planets, and  suggested that stellar abundances
can be used to add constraints on the composition of orbiting rocky planets.
For our studied stars, we have derived  the dependences of [Mg/Si] on [Fe/H]
and [Fe/Si] (see Figs. \ref{mgsife}, \ref{mgsi}).
In the plot of [Mg/Si] versus [Fe/H], the planet-hosting stars 
and the stars without planets have a similar scatter, namely 0.068 and 0.079,
 respectively.
 Unfortunately, in our sample there are no stars with low-mass planets close to 
the Earth-size ones. Nevertheless, the star with the smallest [Mg/Si] ratio 
hosts the planet (HD 156668) with the lowest mass in this sample (0.01M$_J$).

\subsection{Iron-peak elements}
For the iron-peak elements manganese and nickel (Figs.\ref{abun1},
\ref{abun2}) we see no significant difference either.
This contradicts  the study \cite{kang:11}, which reported some differences in 
the manganese abundances between the stars with and without planets.

\subsection{Oxygen, sulphur and zinc}
These elements are singled out for special attention as they are assigned
to volatile elements and have been investigated  in numerous studies
\citep[e.g.][etc]{gonzalez:97, smith:01,melendez:09}.
Higher abundance of volatile elements compared with the solar-analogue stars and 
solar twins with giant planets can be a sign of the presence of Earth-like planets.
In a study of the difference between the solar photospheric abundance and elemental 
abundance in B stars and nearby F-G dwarfs,
\cite{gonzalez:97} plotted solar and stellar abundances against
the condensation temperature $T_{cond}$. The condensation temperature of a
given element is the temperature in a gaseous environment at which half of
the element's atoms condense out of the gas plane, usually on to grains.
The gas-phase abundances in the interstellar medium correlate strongly with  $T_{cond}$ so
that the elements with higher  $T_{cond}$ have the lowest gas-phase
abundances. An increase in the relative content of volatile elements
as compared with the solar-analog stars and solar twins with giant
planets can be indicative of the presence of Earth-like planets.
A comparison of the behavior of these elements depending on $T_{cond}$
in the stars with and without planets will be presented below (Sect.4).
A direct comparison of the abundances of the indicated elements with
the metallicity does not show any significant
differences (see Fig.\ref{abun1}).

\subsection{Neutron-capture elements}
The neutron-capture elements show no significant differences in trends;
the observed differences are within the errors (Fig.\ref{abun2}).
However, there is one exception, which is a surprise, namely barium.
This element shows a markedly different behaviour in the stars with and without planets.
However, the results obtained require further verification. 

\begin{figure}
\begin{tabular}{c}
\includegraphics[width=8cm]{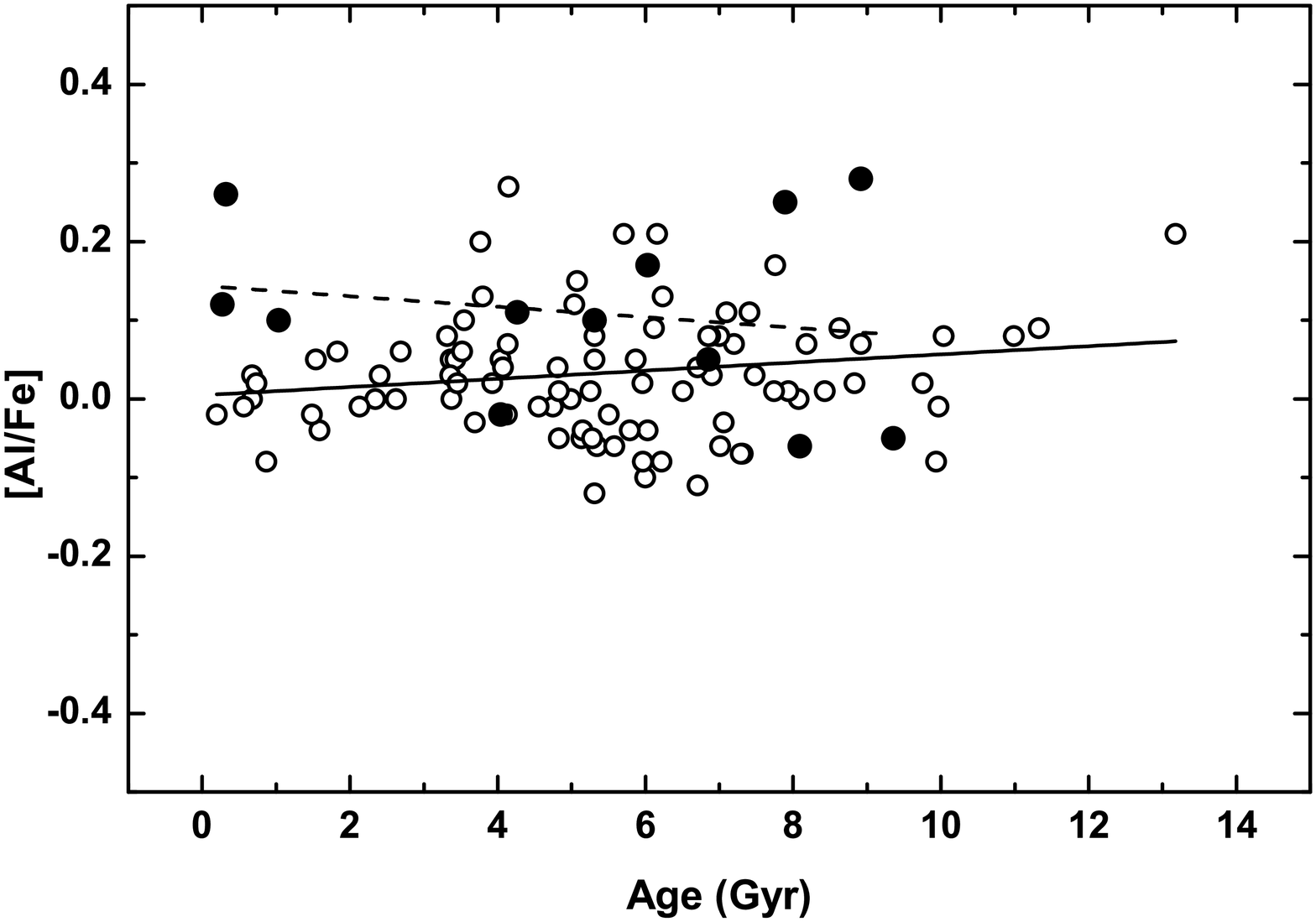}\\
\end{tabular}
\caption{Dependence of [Al/Fe] on age for our target stars. Designations are as in Fig. \ref{fe_teff}. }
\label{al_age}
\end{figure}

\begin{figure}
\begin{tabular}{c}
\includegraphics[width=8cm]{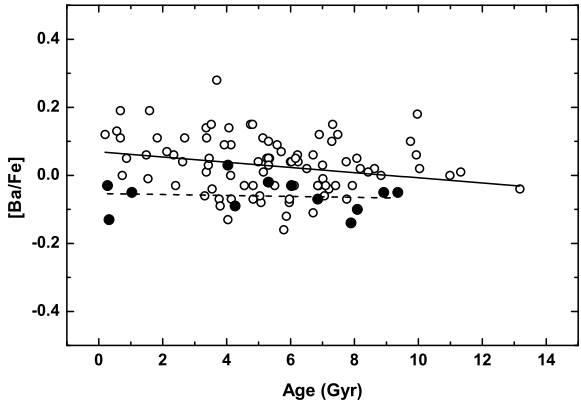}\\
\end{tabular}
\caption{Dependence of [Ba/Fe] on age for our target stars. Designations are as in Fig. \ref{fe_teff}.}
\label{ba_age}
\end{figure}

\section{The dependence of the elemental abundances on the condensation temperature $T_{cond}$ }
\label{sec: Tcond_abun}

\cite{melendez:09} found that volatile elements
(having low $T_{cond}$ ) are more abundant in the Sun relative to
in Solar twins while elements that easily form dust (elements with high
$T_{cond}$, i.e., refractories) are underabundant. Previous attempts
to trace $T_{cond}$ - dependent abundance differences for stars known
to host giant planets have not convincingly demonstrated significant 
differences \citep{udry:07}.
\cite{gonzalez:13} revisit the sample of solar analogs, in particular,
eight stars hosting super-Earth-like planets. Only  four of them display
clear increasing abundance trends with $T_{cond}$. The authors suggested
that there is 
no clear evidence supporting the hypothesis that the volatile-to-refractory
abundance ratio is related to the presence of rocky planets. 
\cite{adibekyan:14, adibekyan:15a} found that the chemical
peculiarities (i.e. small refractory-to-volatile ratio) of planet-hosting
stars is probably
a consequence of their older age and their inner Galaxy origin. 
They conclude that the
stellar age and probably the Galactic birthplace are key to establish the
abundances of some specific elements. 
\cite{nissen:15} indicated that the dependence of [El/Fe] on 
stellar age and the [Ni/Fe]--[Na/Fe] variations complicate the use of
the [El/Fe]--$T_{cond}$ relation
as a possible signature for the existence of terrestrial planets around
stars. 
He assumed that the age trends for the various abundance ratios provide
new constraints on 
supernovae yields and Galactic chemical evolution.
Note that \cite{spina:16} claimed  that it is possible to disentangle 
the signature of planets and chemical evolution.
We also have made an attempt to compare the volatile and refractory element abundances
in solar twins and planet-hosting stars with the temperatures in the range
5750--5900 K.

\subsection{16 Cyg A (HD186408) and 16 Cyg B (HD186427)}

The star 16 Cyg is a well-known and well-studied binary system, with
one component having one planet
(16 Cyg B) and the other one (16 Cyg A)  without a detected planet
\cite[e.g.][]{friel:93, cochran:97, deliyan:00, tucci:14}. 
Both stars are included in our stellar sample. A comparison of stellar parameters
with those of other authors \citep{deliyan:00, tucci:14} is presented in Table \ref{16_comp}.

\begin{table*}
\caption{Parameter comparison for the stars 16 Cyg A and 16 Cyg B . }
\label{16_comp}
\begin{tabular}{ccllll}
\hline
Star     & HD     &     \Teff\ (K)&   \logg     &     [Fe/H]      &        references     \\
\hline
16 Cyg A & 186408 & 5803$\pm$4  & 4.20$\pm$0.15  &  0.09$\pm$0.08   &          our          \\
   --    &   --   & 5795$\pm$20 & 4.30$\pm$0.06 &  0.04$\pm$0.02   &   \cite{deliyan:00}   \\
   --    &   --   & 5830$\pm$7  & 4.30$\pm$0.02 &  0.101$\pm$0.008 &   \cite{tucci:14}     \\
16 Cyg B & 186427 & 5752$\pm$4  & 4.20$\pm$0.15  &  0.02$\pm$0.08   &          our          \\
   --    &  --    & 5760$\pm$20 & 4.40$\pm$0.06 &  0.06$\pm$0.02   &    \cite{deliyan:00}  \\
   --    &  --    & 5751$\pm$6  & 4.35$\pm$0.02 &  0.054$\pm$0.008 &     \cite{tucci:14}   \\
\hline
\end{tabular}
\end{table*}

We obtained a good agreement between these independent determinations.
16 Cyg B is cooler than 16 Cyg A, and 
its [Fe/H] is slightly lower than that of 16 Cyg A.

The lithium abundance is different in these two stars:
we obtained the upper limit of lithium abundance for 16 Cyg B as log A(Li) $<$ 0.80
and the lithium abundance for 16 Cyg A is log A(Li) = 1.45. 
The lithium abundance of the star with the detected planet is lower than that of the star
without a planet.
This is in agreement with the results of \cite{deliyan:00}.
The lithium abundance difference may be due to the accretion of planetary material by
the A component, as proposed by \citep{deliyan:00}.
It should be noted that there 
is no consensus on the results  for this  system
\citep{laws:01, takeda:05, schuler:11, tucci:14}. 

\begin{figure}
\begin{tabular}{c}
\includegraphics[width=8cm]{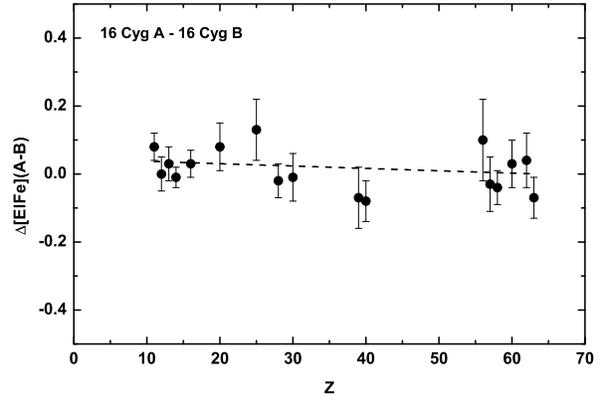}\\
\end{tabular}
\caption{Dependence of $\Delta$[El/Fe]  on  Z for 16 Cyg A and B.}
\label{16AB_z}
\end{figure}

\begin{figure}
\begin{tabular}{c}
\includegraphics[width=8cm]{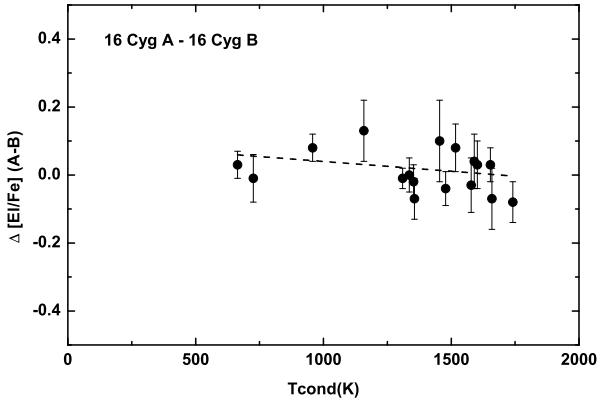}\\
\end{tabular}
\caption{Dependence of  $\Delta$[El/Fe]  on $T_{cond}$ for 16 Cyg A and B. }
\label{16AB_tc}
\end{figure}

The trend for the abundance difference $\Delta$[El/Fe] with the atomic number Z
and with the condensation temperature $T_{cond}$ for these two stars is
presented in Figs. \ref{16AB_z}, \ref{16AB_tc}. As can be seen from
these figures, there is no apparent trend of the difference in elemental
abundances with either atomic number or the condensation temperature for
these two stars. Hence, we cannot confirm the results obtained in \citep{tucci:14}
for this system. For 16 Cyg A and B, these authors found that all elements
showed abundance differences between the binary components. However, while
the difference for volatile elements is about 0.03 dex, the
refractory abundance difference is larger, and the latter showed a
trend with condensation temperature, which could be interpreted as the
signature of a rocky accretion core in the giant planet 16 Cyg Bb.
However, it should be note that the quality of our data is not very high; hence,  
we cannot provide better constraints than those given in our previous works.

\begin{figure}
\begin{tabular}{c}
\includegraphics[width=8cm]{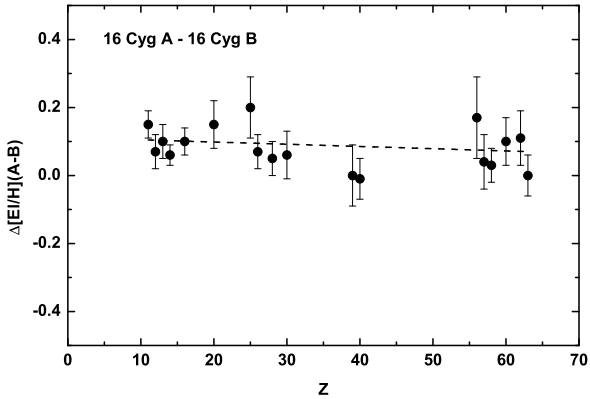}\\
\end{tabular}
\caption{Dependence of $\Delta$[El/H]  on  Z for 16 Cyg A and B.}
\label{16AB_z_}
\end{figure}

\begin{figure}
\begin{tabular}{c}
\includegraphics[width=8cm]{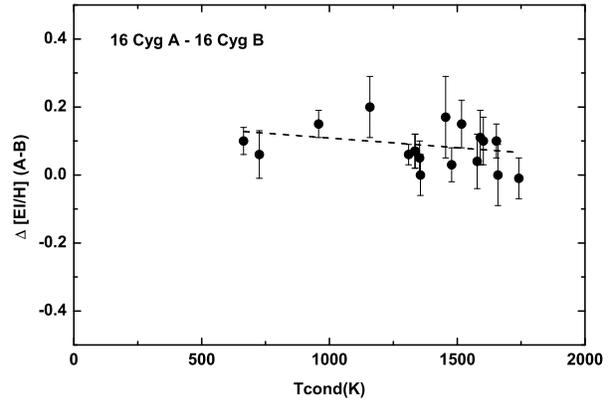}\\
\end{tabular}
\caption{Dependence of  $\Delta$[El/H]  on $T_{cond}$ for 16 Cyg A and B. }
\label{16AB_tc_}
\end{figure}

We also examined the trends for abundance differences relative
to hydrogen $\Delta$[El/H] with the atomic number Z and with condensation
temperature $T_{cond}$ for these two stars (Figs. \ref{16AB_z_}, \ref{16AB_tc_}).
In this case, using [El/H], we obtained a slope similar to the one for
[El/Fe] (--5.71E-5 with standard error SE = 5.42E-5).
However, the star without a planet, 16 Cyg A, exhibits high elemental
abundances compared with the planet-hosting star 16 Cyg B, with
average values $<$[El/H](A--B)$>$ = 0.08 $\pm$0.02 and
$<$[El/Fe](A--B)$>$ = 0.01 $\pm$0.02. The overabundance relative to
hydrogen is slightly higher than the determination  errors
(0.07 dex of that surplus is a result of the difference in the iron abundance
obtained for the two stars, [Fe/H]A = 0.09 and [Fe/H]B = 0.02).

\subsection{HD 38858, 95128, 189733, 217014}

\begin{figure*}
\begin{tabular}{cc}
\includegraphics[width=6.4cm]{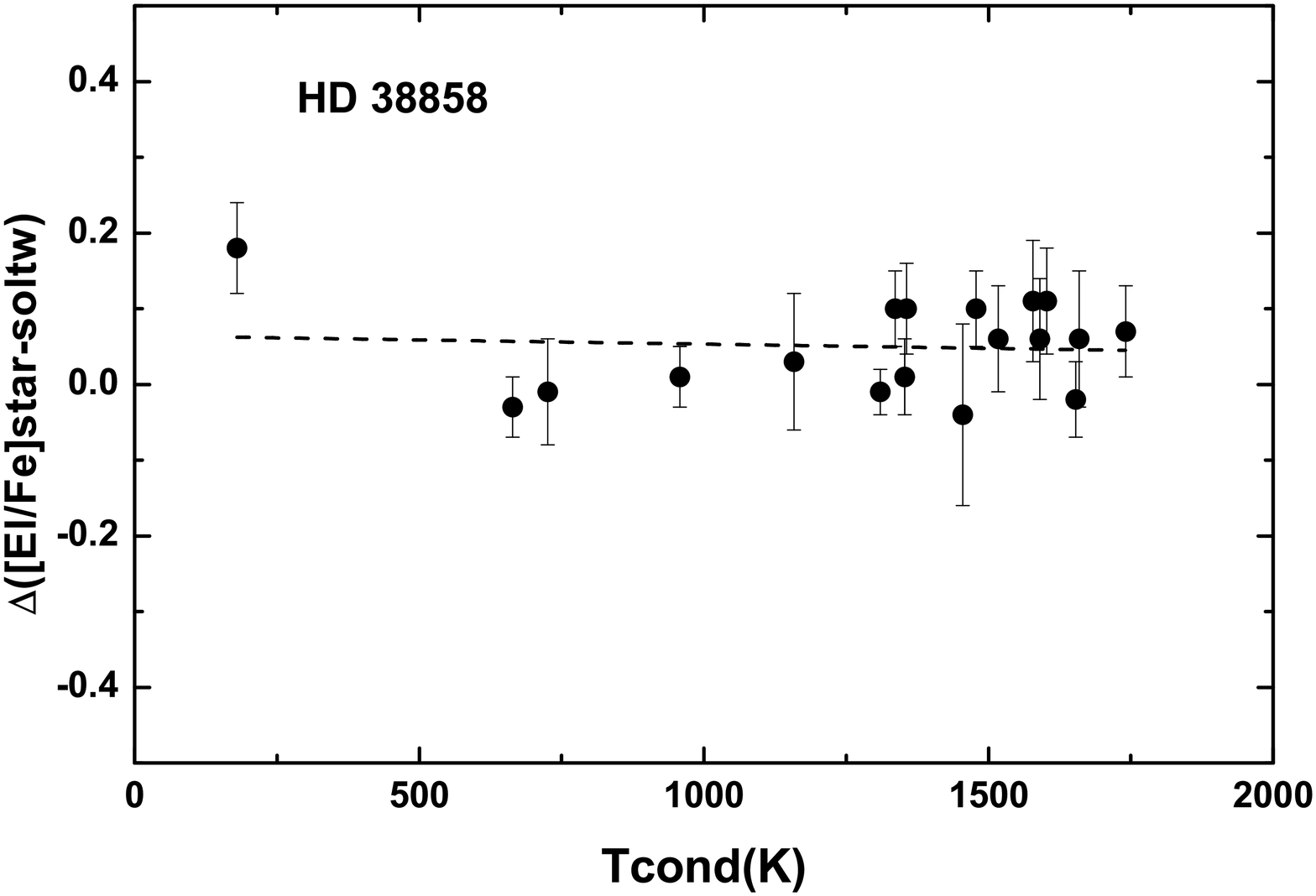}  &\includegraphics[width=6.4cm]{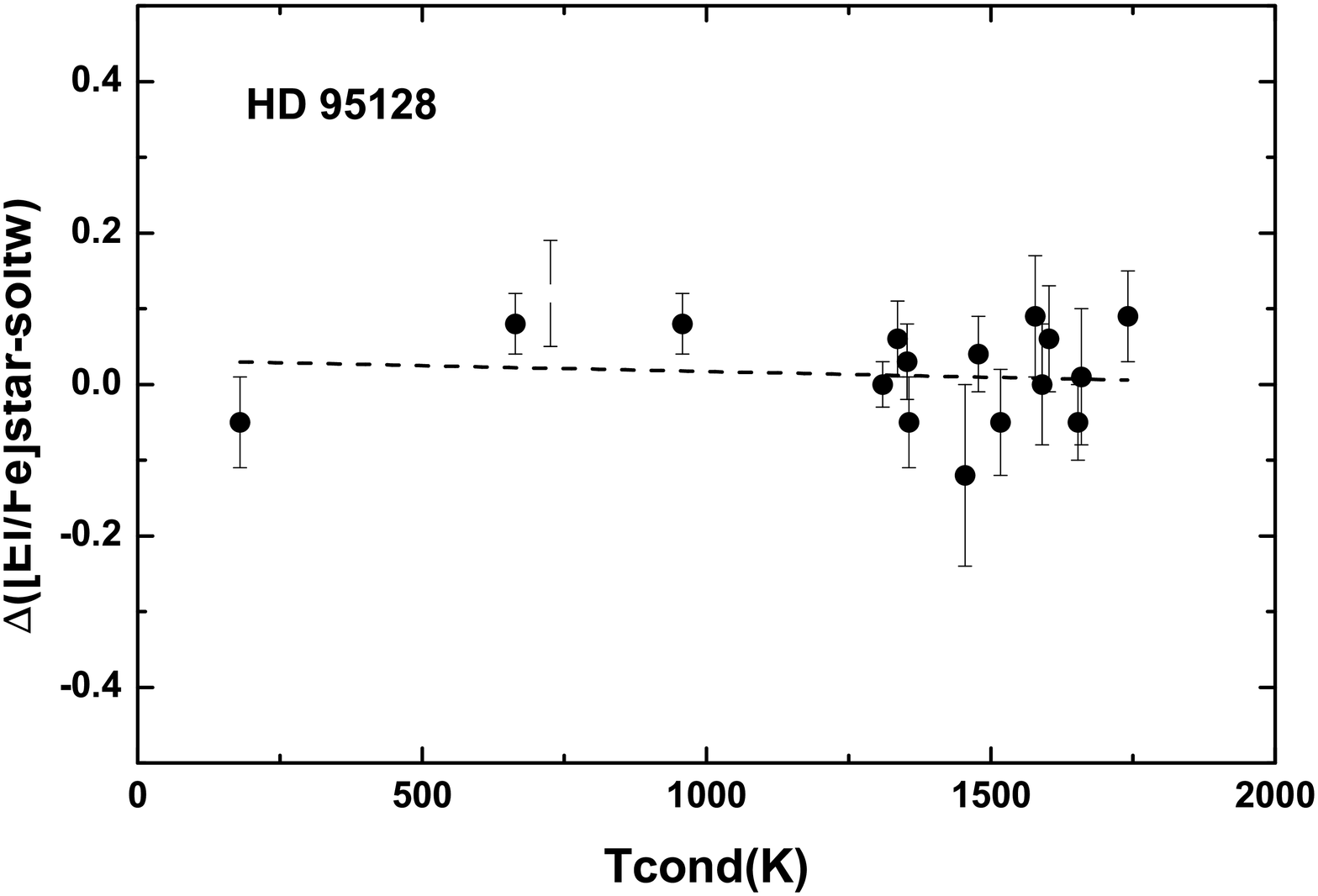}  \\
\includegraphics[width=6.4cm]{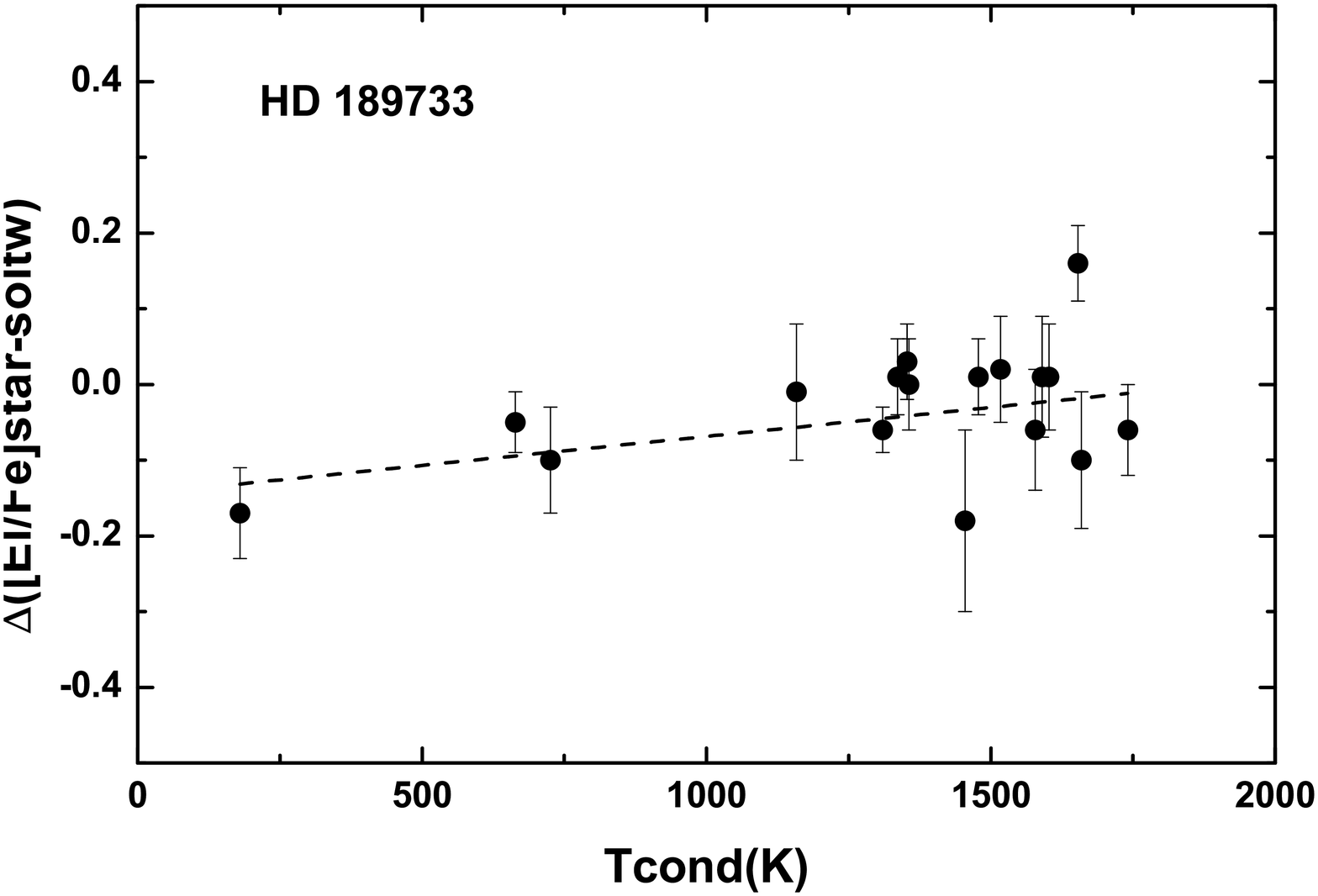} &\includegraphics[width=6.4cm]{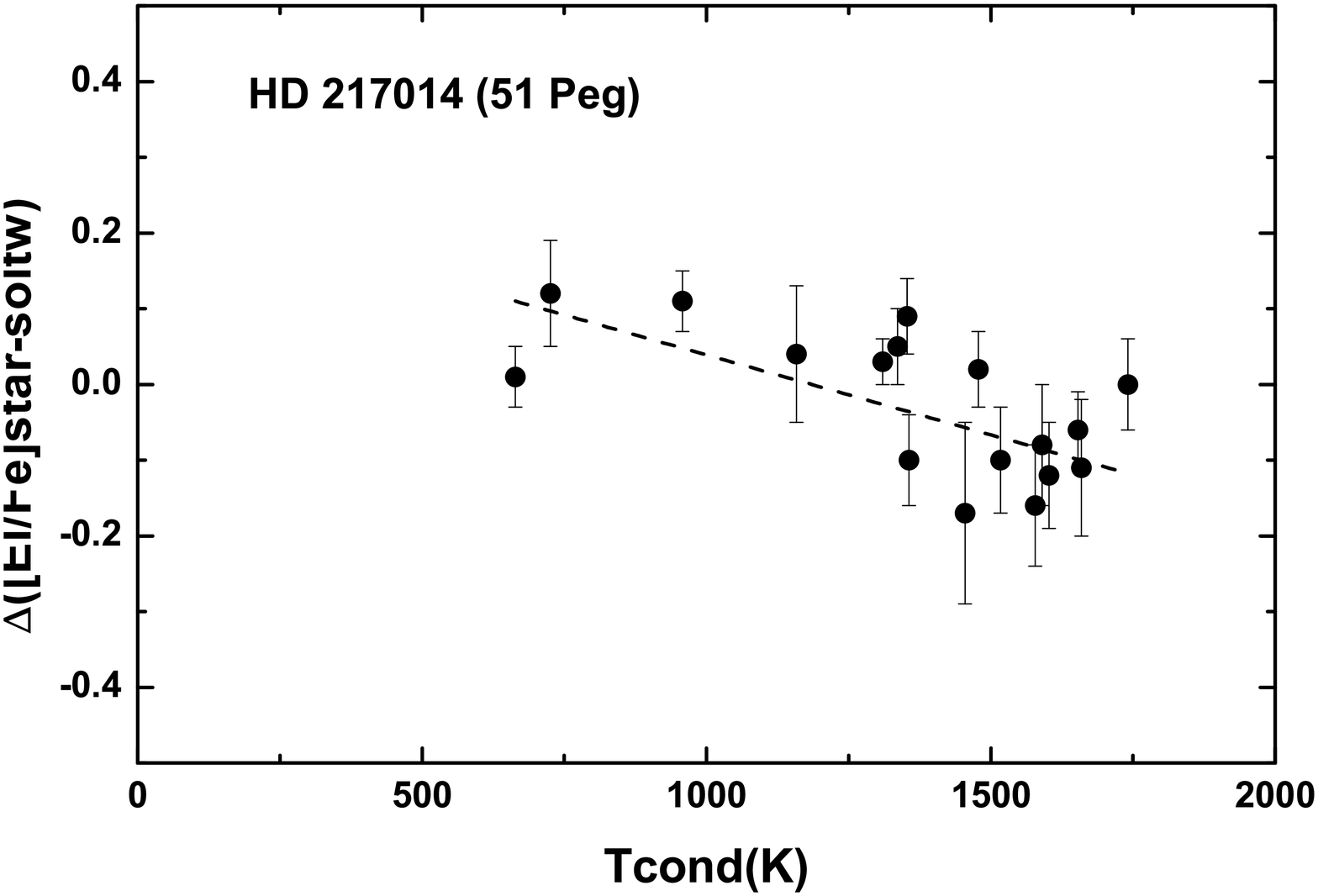}\\
\end{tabular}
\caption{Dependences of $\Delta$[El/Fe](star-solar twins) on  $T_{cond}$ for planet-hosting stars similar to the Sun.}
\label{t_cond}
\end{figure*}

Fig. \ref{t_cond}  shows the dependence of the difference of the elemental
abundances in planet-hosting stars between the mean abundance of
solar twin stars and $T_{cond}$ for four stars whose parameters
are similar to solar (HD 38858, 95128, 189733, 217014=51 Peg).
These stars have massive planets whose masses range from 0.1 M$_J$ to 2M$_J$.  
It can be seen that the dependence for HD217014 (51 Peg) (M $\sim$ 0.5M$_J$) exhibits
a noticeable slope, while the correlations for HD 38858 (M $\sim$ 0.1M$_J$)
and HD 95128 (with two known planets with masses 1.6 M$_J$ and 0.5 M$_J$) do not
exhibit any slope. Volatile elements (low $T_{cond}$) are more abundant in
51 Peg relative to solar twins, while refractory elements (high 
$T_{cond}$) (which may form rocky cores of massive planets) are underabundant.
HD189733 also shows a slope of $\Delta$[El/Fe](star--solar twins) dependence on
$T_{cond}$, but with the opposite sign.
Table \ref{slope_er} presents the linear fitting parameters for
 $\Delta$[El/Fe] versus $T_{cond}$ for the above-mentioned stars.

\begin{table}
\caption{Linear fitting parameters (slope and standard error)
for $\Delta$[El/Fe] versus $T_{cond}$ . }
\label{slope_er}
\begin{tabular}{ccrr}
\hline
Star           & HD            &    slope  &  st.error \\
\hline
16 Cyg (A--B)  & 186408/186427 & --5.77E-5 & 5.59E-5   \\
   --          &  38858        & --1.11E-6 & 3.76E-5   \\
 47 UMa        &  95128        & --1.54E-5 & 4.53E-5   \\
   --          & 189733        &   7.66E-5 & 4.95E-5   \\
 51 Peg        & 217014        & --2.11E-4 & 6.30E-5   \\
\hline
\end{tabular}
\end{table}

There are several possible explanations for the results obtained:
for each star investigated for the presence of planets,
an individual approach for the chemical composition analysis is required, 
as stars were formed in different areas of the Galactic disc with different
pre-stellar elemental 
abundances (as indicated by \cite{adibekyan:14}); or because it is evidence
of the supernovae yields and Galactic chemical evolution \citep{nissen:15}.

\section{Connection with chemical enrichment of different galactic substructures}
\label{sec: connect_struct}

All of the presented comparisons were made for the stars in the thin
disc of the Galaxy, as the 14 selected planet-hosting stars are thin-disc stars. 
The association of the stars in our data base with Galactic
substructures (thin and thick discs and halo) was determined earlier
by us using kinematic criteria. 
The trends of [El/Fe] versus [Fe/H] are typical for thin-disc stars.
Thus, in particular, an increasing of the magnesium abundance with  metallicity was observed 
for the thin-disc stars, but to a lesser degree than  for those of the thick disc.
For Al and Ba, we investigated the dependences [Al/Fe] and [Ba/Fe] on age (Figs. \ref{al_age}, 
\ref{ba_age}).
We can see that the difference in the Al and Ba behaviour with age is very small, and 
is similar to that for metallicity. 

The resulting difference in the barium behavior in the stars with and
without planets could not be interpreted owing to the fact that the barium
abundance depends on the age in the Galactic disc, as found
by \cite[e.g.][]{bensby:07}. 
We were able to determine the real age  for only five stars among 14 planet-hosting stars. 
The ages of these stars range from 5 to 9 Gyr.  
The age of the youngest is  5.31 Gyr  and [Ba/Fe] =--0.02,
the age of the oldest is  8.92 Gyr and [Ba/Fe] = --0.05.
We see the same barium deficit for the planet-hosting stars of all ages in our sample. 
Furthermore, our previous studies did not find a clear correlation between the
barium abundance and the age in the thin disc, which is, in our
opinion, a result of the dispersion of metallicity in the thin disc.
Moreover, our previous studies of the barium abundances in open cluster
stars showed wide variation of the barium content from cluster to cluster,
as well as higher barium abundances in young clusters 
\citep[][]{mishenina:13, mishenina:15b}.

\section{Results and discussion}
\label{sec: conclus}

Using homogeneous spectral data and techniques for determination of
parameters and abundances of a number of elements, we compared
results obtained for stars with and without planets. We have
examined a total of about 200 stars, including  14 planet-hosting stars.
Our main findings are as follows. \\
-- The lithium abundances in planet-hosting solar-analog stars (i.e. stars
with parameters very similar to the solar) in our small sample were lower than
those in the stars for which no planetary systems had been discovered.
The lithium abundance does not exceed 1.7 according to the scale
where hydrogen is 12.0, except one star, namely HD 9826. \\
-- For the binary star 16 Cyg, the star with a planet (16 Cyg B)
has a lower lithium abundance than its companion without a detected planets.\\
-- No significant differences, exceeding determination errors, for the abundances of other
elements were found between stars with and without planets, with possible exception 
 for aluminum and barium abundances. \\
-- No statistically significant dependences of abundance differences in [El/Fe] and [El/H] with the condensation temperature $T_{cond}$ for the two stars in the 16 Cyg binary system were found; 
it was not possible to determine the slope in 16 Cyg because of the large error bars.\\
-- The abundance difference $<$[El/H](A--B)$>$ = 0.08 $\pm$0.02 for the two stars
shows a slight surplus, the origin of which is not clear.\\
-- A slight excess of volatile elements and a deficit of refractories as compared with
solar twins were obtained for 51 Peg (HD 217014).\\

The chemical composition of stars as well as atmospheric parameters 
are important diagnostic tools in studies of stellar evolution,
nucleosynthesis computations, and studies of various physical processes both inside 
stars and on their surfaces, etc. The application of chemical composition
analysis always requires careful consideration of the origin of the chemical composition 
and the mechanisms of its change. As noted above,
much attention has been paid recently to the investigation of any correlation
between the abundance of one or another element and the presence of planetary
systems around stars. 
 
{\it Metallicity.} It is now clear that massive planets are observed preferentially 
around stars with solar or higher
metallicities \citep[][]{fischer:05, sousa:08}, while Earth-like planets 
can be hosted by stars with different
metallicities \citep{sousa:08, adibekyan:12a, adibekyan:12b, buchhave:12}.
Our current sample contains several stars hosting Jupiter-like planets. Among them
there is the only star (HD 154345) that hosts a single planet with
mass M = 1M$_J$.
The most metal-poor planet-hosting star in our sample has the metallicity of 
 $\sim$ --0.3 dex and  host a low-mass planet with a 
mass of $\sim$ 6.5 $M_{\oplus}$ (HD 97658). The requirement of
high metallicity may be a significant  factor for the presence
of massive planets, but this is not the case for less massive planets.

{\it Lithium.} This element is exposed to the effects of a number of 
factors to the larger extent than iron, which results in its abundance variations. 
In general, the lithium abundance
is found to be lower in planet-hosting stars
\citep[][etc.]{gonzalez:00, gonzalez:08, gonzalez:10, israelian:04,
israelian:09, delgado:14,  figueira:14, delgado:15}.
The depletion of lithium owing to convection is observed in the
stars with low temperatures (\Teff\ is lower that 5600 K)
\citep[e.g.][]{randich:10}. With temperatures close to 
solar and higher, the change in the lithium abundance
is caused by the chromospheric activity \citep[e.g.][]{tati:07}
and (or) stellar rotation \cite{charb:92}. It should also be  note that 
many authors  \cite[e.g.][etc]{baumann:10, carlos:16} have 
found strong evidence for increasing lithium depletion with age. 
The low lithium 
abundance determinations obtained by us for stars with
massive planetary systems sustain the hypothesis regarding 
lithium depletion during the formation of stars, but do not
preclude the possibility of specific enrichment of the
pre-stellar cloud from which the star was formed. 
The number of stars in our sample is not high enough to assert the presence of 
dependence of the lithium abundance on the age of stars. 

{\it Volatile and refractory elements.} The abundance determinations for
elements with various condensation temperatures $T_{cond}$ in
parent stars, which could be depleted during the formation of planets, means that we can
suggest the possibility of the formation of rocky Earth-like
planets or of rocky cores of massive planets
\citep[e.g.][etc]{gonzalez:97, smith:01,melendez:09}.
Our data showed the presence of a trend (a slope of the plotted
dependence) for the relative elemental abundances (star -- mean
elementary abundances in solar twins) with $T_{cond}$ only
for one star among four. Thus, for the star HD 217014 (51 Peg)
we found a slight excess of volatile elements and a deficit of
refractories as compared with those obtained for the solar twins.
It should be noted that 51 Peg has an age that is similar to the solar
value \citep[e.g.][]{israelian:04}. The absence of such correlation
for other three stars just sustains the assumption that the stellar
chemical composition can be reckoned as better reflection of the
chemical and dynamic evolution of the Galaxy, in particular the
enrichment of the start's place of origin by supernovae of
different types at the given time intervals
\citep{adibekyan:14, nissen:15}.

{\it Other elements}. \cite{kang:11} claimed different manganese abundance
between the star with and without planets. Our studies have not confirmed
any difference in the Mn abundances for our sample stars with and
without planets \citep{kang:11}; however, we found that aluminium
is more abundant while barium is less abundant in the stars with
planetary systems. This fact requires further verification and confirmation. 

Using our data base, we can corroborate independently that the chemical
composition is undoubtedly an important, albeit an auxiliary, factor
to be taken into account when studying the presence and formation of
planetary systems. On the one hand statistical approaches (based
on relative studies) for large samples of stars are required; on
the other hand it is necessary to investigate each single star and
each chemical element individually.

\section*{Acknowledgements}
This paper is based on the observations collected at OHP observatory, France.
We gratefully acknowledge the anonymous referee for the constructive comments and suggestions.
TM  thanks for the support from the Swiss National Science
Foundation, project SCOPES No. IZ73Z0$_{}$152485. 
VZhA acknowledges the support of the Funda\c{c}\~ao para a Ci\^encia e Tecnologia (FCT - Portugal) 
in the form of the grant SFRH/BPD/70574/2010. VZhA was also supported by FCT through the research 
grants (ref. PTDC/FIS-AST/7073/2014 and ref. PTDC/FIS-AST/1526/2014) through national funds and 
by FEDER through COMPETE2020 (ref. POCI-01-0145-FEDER-016880 and ref. POCI-01-0145-FEDER-016886).



\clearpage

{\bf Appendix}
\begin{table*}
{\bf Table A1. Comparison of atmospheric
parameters for each hosting planet stars
in our sample with the data of other authors}\\
\begin{tabular}{rlcrl}
\hline
HD&\Teff (K) &\logg &[Fe/H]& sources  \\
\hline
\hline
3651&5277& 4.50  &0.15  &  our       \\
    &5182& 4.30  &0.12  & \cite{santos:13}   \\
    &5303& 4.56  &0.18  &  \cite{ramirez:12} \\
    &5252& 4.32  &0.19  &  \cite{ghezzi:10}   \\
    &5196& 4.49  &0.20   & \cite{mald:12}   \\
7924&5165   & 4.40   &--0.22  &  our  \\
    &5133   & 4.46   &--0.22  &  \cite{santos:13} \\
\hline
\hline
\end{tabular}
\end{table*}

\clearpage

\begin{table*}
{\bf Table A2. Abundance of hosting planet stars}\\
\tiny
\begin{tabular}{rrlrrrrrrrrrrrrr}
\hline
 HD    &\Teff (K) & logg  & [Fe/H]  &  Li   &li$_{up}$& [O/Fe]  & [Na/Fe]& [Al/Fe] & [Mg/Fe] & [S/Fe]  & [Si/Fe] & [Ca/Fe]  & [Mn/Fe]    &[Ni/Fe] &  [Zn/Fe]  \\
\hline
\hline
3651   & 5277  &  4.5  &   0.15  &  --   &  --0.52 &  --0.32 &   --   &   0.25  &  --0.02 &  --0.11 &   0.05  &   0.01   &   0.07     &  0.11  &   --0.01  \\
7924   & 5165  &  4.4  &  --0.22 &  --   &  --0.30 &   0.10  &   --   &   0.10  &   0.18  &  --0.19 &   0.09  &   0.07   &  --0.03    &  0.01  &   --0.11  \\
9826   & 6074  &  4.0  &   0.10  &  2.23 &   --    &  --0.15 &  --0.05&   0.10  &   0.09  &  --0.12 &   0.15  &   0.04   &  --0.07    & --0.01 &    0.08   \\
38858  & 5776  &  4.3  &  --0.23 &  1.6  &   --    &   0.16  &  --0.06&  --0.02 &   0.11  &  --0.11 &   0.04  &   0.10   &  --0.01    &  0.01  &   --0.07  \\
87883  & 5015  &  4.4  &   0.00  &  --   &   --    &   0.08  &   --   &   0.28  &   0.03  &   0.04  &   0.12  &   0.07   &   --       &  0.10  &    0.08   \\
95128  & 5887  &  4.3  &   0.01  &  1.7  &   --    &  --0.07 &   0.01 &  --0.05 &   0.07  &   0.00  &   0.05  &  --0.01  &   --       &  0.03  &    0.06   \\
97658  & 5136  &  4.5  &  --0.32 &  --   &  --0.3  &   0.25  &   --   &   0.12  &   0.08  &   0.11  &   0.13  &   0.01   &   --       &  0.03  &   --0.01  \\
\hline
\hline
\end{tabular}
\end{table*}

\begin{table*}
{\bf Table A2  (continued)}\\
\label{}
\begin{tabular}{rrlrrrrrrrrr}
\hline
 HD    & \Teff (K)& \logg & [Fe/H]  &[Y/Fe]  & [Zr/Fe]   &[Ba/Fe] &[La/Fe]  &[Ce/Fe]  &[Nd/Fe] &  [Sm/Fe]   & [Eu/Fe]  \\
\hline
\hline
3651   & 5277  &  4.5  &   0.15  &  --0.12&   --0.11  & --0.14 &  --0.23 &  --0.12 & --0.09 &    --0.18  &   --0.08 \\
7924   & 5165  &  4.4  &  --0.22 &   0.00 &    0.03   & --0.05 &  --0.01 &   0.10  &  0.12  &     0.06   &    0.04  \\
9826   & 6074  &  4.0  &   0.10  &  --0.11&    --     & --0.02 &  --0.27 &  --0.08 &  --    &    --0.06  &    --    \\
38858  & 5776  &  4.3  &  --0.23 &   0.01 &   --0.02  &  0.03  &  --0.02 &   0.05  &  0.10  &     0.01   &    0.15  \\
87883  & 5015  &  4.4  &   0     &  --0.05&   --0.15  & --0.05 &  --0.08 &  --0.01 &  0.08  &    --0.05  &    0.02  \\
95128  & 5887  &  4.3  &   0.01  &  --0.04&    0.00   & --0.05 &  --0.04 &  --0.01 &  0.05  &    --0.05  &    0.00  \\
97658  & 5136  &  4.5  &  --0.32 &  --0.12&    0.01   & --0.03 &  --0.09 &  --0.04 &  0.06  &     0.05   &    0.19  \\
128311 & 4960  &  4.4  &   0.03  &  --0.07&    0.00   & --0.03 &  --0.17 &  --0.04 &  0.02  &    --0.08  &    0.04  \\
145675 & 5406  &  4.5  &   0.32  &  --0.05&   --0.10  & --0.09 &  --0.16 &   0.03  &  0.00  &    --0.07  &   --0.03 \\
154345 & 5503  &  4.3  &  --0.21 &   0.10 &   --0.04  & --0.05 &  --0.19 &   0.07  &  0.04  &     0.02   &    0.15  \\
156668 & 4850  &  4.2  &  --0.07 &  --0.12&   --0.09  & --0.13 &  --0.15 &  --0.07 &  0.05  &    --0.02  &    0.05  \\
186427 & 5752  &  4.2  &   0.02  &  --0.04&   --0.03  & --0.07 &  --0.05 &   0.03  & --0.07 &    --0.11  &    0.02  \\
189733 & 5818  &  4.3  &  --0.03 &  --0.15&   --0.15  & --0.11 &  --0.19 &  --0.04 &  0.00  &    --0.04  &    0.05  \\
217014 & 5778  &  4.2  &   0.14  &  --0.16&   --0.09  & --0.10 &  --0.29 &  --0.03 & --0.13 &    --0.13  &   --0.05 \\
\hline
\hline
\end{tabular}
\end{table*}

\begin{table*}
{\bf Table A3. Abundance of stars without planets (Li-Zn).
Solar-analogue stars are marked by asterisks.}
\tiny
\begin{tabular}{rrlrrrrrrrrrrrrr}
\hline
\hline
HD  &    \Teff (K)& \logg &  [Fe/H] &   Li   &li$_{up}$& [O/Fe]   & [Na/Fe]  &  [Al/Fe]  &  [Mg/Fe]   &   [S/Fe]  &  [Si/Fe]  & [Ca/Fe] &[Mn/Fe] &  [Ni/Fe] &  [Zn/Fe]   \\
\hline
   166 & 5514  &  4.6  &   0.16  &  2.24  &  --     &   --0.23 &    --    &    --0.02 &     --0.13 &     --    &    --0.02 &    0.03 &  0.01  &    0.02  &    0.04    \\
  1562 & 5828  &  4.0  &  --0.32 &  1.9   &  --     &    0.16  &    0.04  &     0.13  &      0.18  &     0.01  &     0.07  &    0.23 & --0.04 &   --0.02 &   --0.01   \\
  1835 & 5790  &  4.5  &   0.13  &  2.55  &  --     &   --0.07 &   --0.10 &     0.00  &      0.00  &    --0.04 &     0.07  &    0.02 &  0.07  &    0.03  &    0.05    \\
  3651 & 5277  & 4.5   &   0.15  &  --   &  --0.52  &   --0.32 &   --     &      0.25 &    --0.02  &    --0.11 &      0.05 &   0.01  &  0.07  &    0.11  &   --0.01   \\
  4256 & 5020  & 4.3   &   0.08  &  --   &  --0.40  &   --0.25 &   --     &      0.38 &     0.08   &     --    &      0.17 &   0.12  &  0.06  &    0.16  &    0.07    \\
  4307 & 5889  &  4.0  &  --0.18 &  2.48  &  --     &    0.17  &   --0.10 &     0.06  &      0.06  &    --0.08 &     0.05  &    0.06 & --0.06 &    0     &    0.02    \\
  4614 & 5965  &  4.4  &  --0.24 &  2.17  &  --     &    0.33  &   --0.09 &     0.09  &      0.10  &    --0.03 &     0.09  &    0.02 & --0.04 &    0.02  &    0.10    \\
\hline
\hline
\end{tabular}
\end{table*}

\begin{table*}
{\bf Table A3  (continued)}\\
\tiny
\begin{tabular}{rrrrrrrrrrrrrrrr}
\hline
\hline
HD  &    \Teff (K)& \logg & [Fe/H]  &   Li   &li$_{up}$&   [O/Fe] & [Na/Fe]  &  [Al/Fe]  &  [Mg/Fe]   &   [S/Fe]  & [Si/Fe]   & [Ca/Fe] &[Mn/Fe] &  [Ni/Fe] &    [Zn/Fe]    \\
\hline
 54371 & 5670  &  4.2  &   0.06  &  1.97  &  --     &   --0.06 &    --    &     0.05  &      0.07  &     --    &    --0.05 &    0.09 &  --    &   --0.02 &   --0.18   \\
 55575 & 5949  &  4.3  &  --0.31 &  1.95  &  --     &          &   --0.12 &     0.08  &      0.21  &     0.00  &     0.10  &    0.05 &  --    &    0.01  &    0.04    \\
 58595 & 5707  &  4.3  &  --0.31 &  1.75  &  --     &    0.16  &    0.00  &     0.05  &      0.07  &    --0.05 &     0.05  &    0.12 & --0.06 &    0.00  &    0.03    \\
 59747 & 5126  & 4.4   &  --0.04 &  1.40 &   --     &    --    &   --     &      0.07 &     0.05   &     --    &      0.05 &   0.06  &  0.05  &   --0.01 &   --0.2    \\
 61606 & 4956  & 4.4   &  --0.12 &  --   &   --     &    0.04  &   0.00   &      0.07 &    --0.10  &     --    &      0.08 &   0.01  &  0.03  &   --0.04 &   --0.05   \\
 62613 & 5541  &  4.4  &  --0.10 &  --    &  --     &    0.08  &   --0.14 &    --0.02 &      0.07  &    --0.06 &     0.04  &   --0.01& --0.03 &    0.02  &   --0.05   \\
 63433*& 5693  &  4.35 &  --0.06 &  2.37  &  --     &    0.10  &    --    &     0.01  &      0.02  &     --    &     0.02  &    0.07 & --0.11 &   --0.05 &   --0.19   \\
\hline
\hline
\end{tabular}
\end{table*}

\begin{table*}
{\bf Table A3  (continued)}\\
\tiny
\begin{tabular}{rrrrrrrrrrrrrrrr}
\hline
\hline
HD  &    \Teff (K)& \logg & [Fe/H]  &   Li   &li$_{up}$&   [O/Fe] & [Na/Fe]  &  [Al/Fe]  &  [Mg/Fe]   &   [S/Fe]  & [Si/Fe]   & [Ca/Fe] &[Mn/Fe] &  [Ni/Fe] &    [Zn/Fe]    \\
\hline
149661 & 5294  & 4.5   &  --0.04 &  --   &   --     &    0.06  &   0.09   &      0.09 &     0.02   &    --0.04 &      0.09 &   0.05  & --0.03 &    0.05  &    0.11    \\
149806 & 5352  & 4.55  &   0.25  &  0.50 &   --     &   --0.25 &   --     &      0.15 &    --0.14  &     --    &      0.07 &  --0.02 &  0.03  &    0.02  &    0.06    \\
151541 & 5368  & 4.2   &  --0.22 &  --   &   --     &    0.23  &   0.00   &      0.13 &     0.05   &     0.06  &      0.03 &   0.07  & --0.06 &   --0.06 &   --0.06   \\
152391 & 5495  &  4.3  &  --0.08 &  1.3   &  --     &    --    &    --    &     --    &      --    &    --0.03 &     --    &    --   &  --    &    --    &     --     \\
154931 & 5910  &  4    &  --0.1  &  2.45  &  --     &    0.04  &    0.01  &     0.06  &      0.14  &    --0.19 &     0.08  &    0.12 &  --    &    0.05  &     0.04   \\
155712 & 4932  & 4.6   &  --0.06 &  --   &  --0.50  &    --    &   --     &      0.16 &    --0.07  &     --    &      0.13 &   --    &  --    &   --0.01 &   --0.10   \\
156668 & 4850  & 4.2   &  --0.07 &  --   &   0.00   &   --0.05 &   --     &      0.26 &     0.08   &     --    &      0.22 &   0.06  & --0.05 &    0.05  &    0.08    \\
\hline
\end{tabular}
\end{table*}

\begin{table*}
{\bf Table A3  (continued: Y--Eu)}\\
\begin{tabular}{rrrrrrrrrrrr}
\hline
HD  &    \Teff (K)& \logg &  [Fe/H] & [Y/Fe]     &  [Zr/Fe]    &  [Ba/Fe]      &    [La/Fe]    &    [Ce/Fe]    &  [Nd/Fe]   &   [Sm/Fe]    &    [Eu/Fe]     \\
\hline
\hline
   166 & 5514  &  4.6  &   0.16  &  --0.05 &   --0.15 &     0.12   &   --0.24   &   --0.12   &   0.00  &   --0.13  &   --0.09    \\
  1562 & 5828  &  4.0  &  --0.32 &  --0.02 &   --0.06 &     0.00   &   --0.02   &    0.07    &   0.08  &    0.07   &    --       \\
  1835 & 5790  &  4.5  &   0.13  &   0.02  &   --0.08 &     0.04   &   --0.22   &    0.01    &  --0.11 &    --     &    0.06     \\
  3651 & 5277  &  4.5  &   0.15  &  --0.12 &  --0.11  &    --0.14  &   --0.23   &   --0.12   &   --0.09&   --0.18  &   --0.08    \\
  4256 & 5020  &  4.3  &   0.08  &  --0.15 &  --0.19  &    --0.16  &   --0.23   &   --0.04   &   --0.04&   --0.03  &             \\
  4307 & 5889  &  4.0  &  --0.18 &  --0.07 &   --0.07 &     0.08   &    0.03    &   --0.04   &   0.05  &    0.06   &    0.12     \\
  4614 & 5965  &  4.4  &  --0.24 &   0.05  &   --0.02 &     0.02   &   --0.06   &   --0.05   &   0.07  &    0.06   &    0.08     \\
\hline
\hline
\end{tabular}
\end{table*}

\begin{table*}
{\bf Table A3  (continued)}\\
\begin{tabular}{rrrrrrrrrrrr}
\hline
HD  &    \Teff (K)& \logg &  [Fe/H] & [Y/Fe]    &  [Zr/Fe]    &  [Ba/Fe]      &    [La/Fe]    &    [Ce/Fe]    &  [Nd/Fe]   &   [Sm/Fe]    &    [Eu/Fe]     \\
\hline
\hline
 54371 & 5670  &  4.2  &   0.06  &  --0.17 &   --0.18 &    --0.01  &   --0.23   &   --0.16   &  --0.13 &   --0.15  &    0.03     \\
 55575 & 5949  &  4.3  &  --0.31 &  --0.07 &    0.04  &     0.02   &    0.06    &   --0.07   &   0.14  &    0.13   &    0.20     \\
 58595 & 5707  &  4.3  &  --0.31 &   0.07  &    0.02  &     0.01   &    0.03    &    0.18    &   0.15  &    0.15   &    0.20     \\
 59747 & 5126  &  4.4  &  --0.04 &  --0.08 &  --0.14  &     0.09   &   --0.15   &    0.04    &    0.04 &   --0.08  &    0.02     \\
 61606 & 4956  &  4.4  &  --0.12 &   0.00  &   --     &     0.02   &    0.03    &    0.19    &    0.15 &    0.20   &    0.13     \\
 62613 & 5541  &  4.4  &  --0.1  &   0.02  &    0.01  &     0.00   &   --0.07   &    0.02    &   0.05  &    0.08   &   --0.06    \\
 63433 & 5693  &  4.35 &  --0.06 &  --0.16 &   --0.18 &     0.02   &   --0.16   &   --0.09   &  --0.02 &   --0.07  &    0.03     \\
\hline
\hline
\end{tabular}
\end{table*}

\begin{table*}
{\bf Table A3  (continued)}\\
\begin{tabular}{rrrrrrrrrrrr}
\hline
HD  &    \Teff (K)& \logg &  [Fe/H] & [Y/Fe]    &  [Zr/Fe]    &  [Ba/Fe]      &    [La/Fe]    &    [Ce/Fe]    &  [Nd/Fe]   &   [Sm/Fe]    &    [Eu/Fe]     \\
\hline
\hline
146233 & 5799  &  4.4  &   0.01  &  --0.02&   --0.05 &     0.01   &   --0.01   &   --0.02   &  --0.01 &   --0.08  &    0.08     \\
149661 & 5294  &  4.5  &  --0.04 &  --0.07 &   0.05   &     0.01   &    0.06    &    0.08    &    0.02 &   --0.09  &    0.03     \\
149806 & 5352  &  4.55 &   0.25  &  --0.18 &  --0.18  &     0.05   &   --0.27   &   --0.08   &   --0.11&   --0.16  &   --0.08    \\
151541 & 5368  &  4.2  &  --0.22 &  --0.14 &   0.02   &    --0.15  &    0.01    &    0.04    &    0.06 &   --0.07  &    0.26     \\
152391 & 5495  &  4.3  &  --0.08 &   --   &    --    &     --     &    --      &    --      &   --    &    --     &    --       \\
154931 & 5910  &  4.0  &  --0.1  &   --   &    --    &     --     &    --      &    --      &   --    &    --     &    --       \\
155712 & 4932  &  4.6  &  --0.06 &  --0.11 &  --0.14  &    --0.12  &   --0.17   &   --0.04   &    0    &   --0.05  &   --0.04    \\
\hline
\hline
\end{tabular}
\end{table*}

\begin{table*}
{\bf Table A3  (continued)}\\
\begin{tabular}{rrrrrrrrrrrr}
\hline
HD  &    \Teff (K)& \logg &  [Fe/H] & [Y/Fe]    &  [Zr/Fe]    &  [Ba/Fe]      &    [La/Fe]    &    [Ce/Fe]    &  [Nd/Fe]   &   [Sm/Fe]    &    [Eu/Fe]     \\
\hline
\hline
219538 & 5114  &  4.4  &  --0.09 &  --0.04 &  --0.05  &    --0.06  &    0.00    &    0.04    &    0.11 &    0.01   &    0.06     \\
219623 & 5949  &  4.2  &   0.04  &  --0.17&   --0.19 &     0.01   &   --0.28   &   --0.2    &  --0.18 &   --0.17  &    0.13     \\
220140 & 5144  &  4.6  &  --0.03 &  --0.15 &  --0.11  &     0.05   &   --0.08   &   --0.16   &   --0.09&   --0.12  &    0.00     \\
220182 & 5364  &  4.5  &  --0.03 &  --0.06 &  --0.20  &     0.07   &   --0.18   &   --0.03   &    0.04 &   --0.06  &    0.10     \\
220221 & 4868  &  4.5  &   0.16  &  --0.02 &  --0.06  &     0.02   &   --0.18   &   --0.06   &   --0.06&   --0.04  &   --0.09    \\
221354 & 5242  &  4.1  &   --    &   --    &   --     &     --     &    --      &    --      &    --   &    --     &    --       \\
221851 & 5184  &  4.4  &  --0.09 &  --0.07 &  --0.13  &     0.02   &   --0.13   &    0.01    &    0.05 &   --0.07  &    0.11     \\
222143 & 5781  &  4.3  &   0.09  &  --0.14&   --0.14 &     0.09   &   --0.23   &   --0.15   &  --0.15 &   --0.17  &   --0.02    \\
224465 & 5745  &  4.5  &   0.08  &  --0.18&   --0.12 &     0.05   &   --0.19   &   --0.13   &  --0.13 &   --0.19  &    0.04     \\
BD+12 063& 4859  &  4.4  &  --0.22 &   0.00  &  --0.02  &    --0.18  &   --0.09   &   --0.03   &   --0.01&   --0.02  &    0.05     \\
\hline
\hline
\end{tabular}
\end{table*}



\end{document}